# SUPERHUMPS IN CATACLYSMIC BINARIES.

## XXIII. V442 OPHIUCHI AND RX J1643.7+3402

JOSEPH PATTERSON,[1,2] WILLIAM H. FENTON,[3] JOHN R. THORSTENSEN,[3]

DAVID A. HARVEY,[4] DAVID R. SKILLMAN,[5] ROBERT E. FRIED,[6] BERTO MONARD,[7]

DARRAGH O'DONOGHUE,[8] EDWARD BESHORE,[9] BRIAN MARTIN,[10] PANOS NIARCHOS,[11]

TONNY VANMUNSTER,[12] JERRY FOOTE,[13] GREG BOLT,[14] ROBERT REA,[15]

LEWIS M. COOK,[16] NEIL BUTTERWORTH,[17] AND MATT WOOD[18]

[1]  Department of Astronomy, Columbia University, 550 West 120th Street, New York, NY 10027; jop@astro.columbia.edu

[2]  Visiting Astronomer, Cerro Tololo Interamerican Observatory, National Optical Astronomy Observatories, which is operated by the Association of Universities for Research in Astronomy, Inc. (AURA) under cooperative agreement with the National Science Foundation

[3]  Department of Physics and Astronomy, Dartmouth College, 6127 Wilder Laboratory, Hanover, NH 03755; William.H.Fenton@dartmouth.edu, thorstensen@dartmouth.edu

[4]  Center for Backyard Astrophysics (West), 1552 West Chapala Drive, Tucson, AZ 85704; dharvery@comsoft-telescope.com

[5]  Center for Backyard Astrophysics (East), 9517 Washington Avenue, Laurel, MD 20723; dskillman@home.com

[6]  Center for Backyard Astrophysics (Flagstaff), Braeside Observatory, Post Office Box 906, Flagstaff, AZ 86002; captain@asu.edu

[7]  Center for Backyard Astrophysics (Pretoria), Post Office Box 70284, Die Wilgers 0041, Pretoria, South Africa; lagmonar@csir.co.za

[8]  South African Astronomical Observatory, Observatory 7935, Cape Town, South Africa; dod@saao.ac.za

[9]  Center for Backyard Astrophysics (Colorado), 14795 East Coachman Drive, Colorado Springs, CO 80908; ebeshore@pointsource.com

[10]  King's University College, Department of Physics, 9125 50th Street, Edmonton, AB  T5H 2M1, Canada; bmartin@kingsu.ab.ca

[11]  University of Athens, Department of Astrophysics, Astronomy, and Mechanics,

Panepistimipolis, GR-157 84, Zografos, Athens, Greece; pniarcho@cc.uoa.gr

[12] Center for Backyard Astrophysics (Belgium), Walhostraat 1A, B-3401 Landen, Belgium; tonny.vanmunster@advalvas.be

[13] Center for Backyard Astrophysics (Utah), 4175 East Red Cliffs Drive, Kanab, UT 84741; jfoote@scopecraft.com

[14] Center for Backyard Astrophysics (Perth), 295 Camberwarra Drive, Craigie, Western Australia 6025, Australia; gbolt@iinet.net.au

[15] Center for Backyard Astrophysics (Nelson), 8 Regent Lane, Richmond, Nelson, New Zealand; reamarsh@ihug.co.nz

[16] Center for Backyard Astrophysics (Concord), 1730 Helix Court, Concord, CA 94518; lcoo@yahoo.com

[17] Center for Backyard Astrophysics (Townsville), 24 Payne Street, Mount Louisa, Queensland 4814, Australia; neilbutt@bigpond.com.au

[18] Florida Institute of Technology, Department of Physical and Space Sciences and SARA Observatory, Melbourne, FL 32901; wood@pss.fit.edu

## ABSTRACT

We report the results of long observing campaigns on two novalike variables: V442 Ophiuchi and RX J1643.7+3402. These stars have high-excitation spectra, complex line profiles signifying mass loss at particular orbital phases, and similar orbital periods (respectively 0.12433 and 0.12056 d). They are well-credentialed members of the SW Sex class of cataclysmic variables. Their light curves are also quite complex. V442 Oph shows periodic signals with periods of 0.12090(8) and 4.37(15) days, and RX J1643.7+3402 shows similar signals at 0.11696(8) d and 4.05(12) d. We interpret these short and long periods respectively as a "negative superhump" and the wobble period of the accretion disk. The superhump could then possibly arise from the heating of the secondary (and structures fixed in the orbital frame) by inner-disk radiation, which reaches the secondary relatively unimpeded since the disk is not coplanar.

At higher frequencies, both stars show another type of variability: quasi-periodic oscillations (QPOs) with a period near 1000 seconds. Underlying these strong signals of low stability may be weak signals of higher stability. Similar QPOs, and negative superhumps, are quite common features in SW Sex stars. Both can in principle be explained by ascribing strong magnetism to the white dwarf member of the binary; and we suggest that SW Sex stars are borderline AM Herculis binaries, usually drowned by a high accretion rate. This would provide an ancestor channel for AM Hers, whose origin is still mysterious.

*Subject headings*: accretion, accretion disks — binaries: close — novae, cataclysmic variables — stars: individual (V442 Ophiuchi) — stars: individual (RX J1643.7+3402)

# 1. INTRODUCTION

Many cataclysmic variables of short orbital period (<4 hr) and high accretion rate ($\dot{M} > 10^{-9}\ M_\odot$/yr) have light curves with prominent humps slightly displaced from $P_{orb}$. These are called "superhumps" since they are particularly characteristic of dwarf novae in superoutburst. Most superhumps occur at periods slightly longer than $P_{orb}$, and have been successfully explained as due to the prograde apsidal precession of an eccentric accretion disk (Whitehurst 1988, Osaki 1989, Lubow 1991). A few, however, occur at periods shorter than $P_{orb}$; these are called negative superhumps, or occasionally *nodal* superhumps to emphasize their probable origin in the retrograde wobble of the disk (Harvey et al. 1995, Patterson et al. 1997, Patterson 1999, Wood et al. 2000, Murray et al. 2002).

V442 Ophiuchi and RX J1643.7+3402 (hereafter RX 1643+34) are bright novalike variables with spectra suggestive of high $\dot{M}$, and fairly short orbital periods (Hoard, Thorstensen, & Szkody 2000; Mickaelian et al. 2002, hereafter M02). So we selected them for long observing campaigns to search for superhumps. Here we report the results. In the process, we may have learned something about the SW Sextantis stars, a mysterious class of cataclysmic variables.

# 2. SPECTROSCOPY OF RX 1643+34

The spectra of RX1643+34 reported by M02 showed a strong blue continuum with weak emission lines of high excitation. From the radial velocities of Hβ they found an orbital period of either 0.1073(2) d or 0.1202(5) d, depending on the (undetermined) daily cycle count. We decided to repeat that study with sufficient baseline and density of observation to yield an accurate and unique period.

Our spectroscopy was carried out in 2002 May and June, using both telescopes at MDM Observatory on Kitt Peak. The procedures essentially followed those described by Thorstensen, Taylor, & Kemp (1998), except that in May we used the 1.3 m telescope and the Ohio State CCD spectrograph. The latter spectra covered 4200–5100 Å, with 0.8 Å pixels and a resolution of 1.9 Å FWHM. The June data consisted of a single orbit with the 2.4 m Hiltner telescope and the modular spectrograph, covering 4300–7500 Å at 2 Å/pixel and 3.5 Å FWHM. Flux standards were observed as appropriate, but the flux calibration in May was only relative.

Figure 1 shows the mean spectrum in June, and Table 1 details the line features. The spectrum resembles that of M02, with a strong blue continuum and weak lines, typical of the most luminous novalike variables.

We measured the apparent radial velocity of Hβ in all spectra, using a gaussian derivative algorithm (Schneider & Young 1980). Figure 2 shows the result of a "residualgram" period search (Thorstensen et al. 1996). Because the observations span a wide range in hour angle, the velocities unambiguously determine an orbital frequency near 8.3 cycles/day. The cycle count between May and June velocities is slightly ambiguous, but the CBA photometry described below in Section 4 confirms the choice which best fits the velocities: $P_{orb}$=0.120560(14) d, in agreement with one of the M02 choices. Figure 3 shows the Hβ velocities folded on $P_{orb}$, and Table 2 gives the parameters of the best-fitting sinusoid.

Because the star is noneclipsing and lacks secondary-star absorptions, we do not have a fiducial marker of absolute binary phase. However, the CBA photometry shows a periodic signal at $P_{orb}$, specifically a double-humped waveform like the classical "ellipsoidal" variation in Roche geometry. In this variation, the shallower of the two minima corresponds to "eclipse phase", i.e., inferior conjunction of the secondary. In Section 4 that ephemeris is found to be

$$\text{Shallower orbital minimum} = \text{HJD } 2,452,383.868(2) + 0.120560(14) \, E, \qquad (1)$$

and we shall adopt that orbital phase convention. If this were fully trustworthy, then the emission lines lag their geometrically required schedule by 72±12°. That would be of interest, since a large phase lag is one of the key indicators for the SW Sex class of cataclysmic variables (Thorstensen et al. 1991). Unfortunately, accretion structures can also produce signals at $P_{orb}$ and $P_{orb}/2$, so this is merely suggestive, not reliable.[1]

Figure 4 shows phase-averaged greyscale images of the spectra from our two observing runs, constructed using procedures described by Taylor et al. (1999). Hα shows an obvious high-velocity component, similar to those seen in LS Peg (Taylor et al. 1999), V533 Her (Thorstensen & Taylor 2000), and V795 Her (Dickinson et al. 1997, Casares et al. 1996). Phase-dependent absorption appears in He I λ4471, λ4921, and λ5015, at orbital phases 0.1<φ<0.6. He II λ4686 shows a narrow component which moves in phase with the Balmer line cores, sharing the same phase lag.

Many of these characteristics are in general accord with the SW Sex class: the high-excitation spectrum, single-peaked emission, apparently large phase lag, 2.9 hr $P_{orb}$, and high-velocity components in the Balmer lines. Phase-dependent He I absorption is also a distinguishing trait, but here appears somewhat different. The founding members of the class show a very distinctive deep absorption quite close to φ=0.5; and this has become a defining characteristic. On the other hand, the founding members were all eclipsers. This poorly understood phenomenon may depend on binary inclination, and perhaps other factors not yet explored (e.g., phase in the 4.05 d cycle discussed below, or in the years-long cycling between negative and positive superhump). So we think the case is overall pretty good for SW Sex membership, although these issues certainly merit further study.

## 3. PHOTOMETRY OF V442 OPHIUCHI

### 3.1 1995 CAMPAIGN

We obtained 165 hours of photometry, spread over 28 nights during 1995 May 20 – July

---

[1] However, CBA photometry shows these double-humped variations in most SW Sex stars; and in every case where it is possible to compare phases against an absolute phase determined from eclipses, they are consistent with ellipsoidal variation (maxima at 0.25 and 0.75). A preliminary version of our light curve atlas can be found at http://cba.phys.columbia.edu/atlas/.

8. Most of the observations consisted of long photoelectric time series in *V* light, acquired with an aperture photometer mounted on the 1.0 m telescope at Cerro Tololo Inter-American Observatory. We used an integration time of 10 s, then subtracted the sky background and used a mean extinction of 0.15 mag/airmass to reduce the time series to counts/integration above the atmosphere. Table 3 summarizes the observations, including those of other stars studied in this paper.

Another 60 hours of coverage were obtained with CCD photometers at the Tucson (35 cm telescope) and Maryland stations of the Center for Backyard Astrophysics. CBA data analysis techniques were discussed by Skillman & Patterson (1993), and yield a time series of differential magnitudes. On several nights there were overlaps with the photoelectric time series, and we used the overlap to convert the differential magnitudes to equivalent count rates in the CTIO time series. The estimated accuracy of the conversion is only ~0.05 mag, but this enabled us to make a smooth splice and extend the time series.

A sampler of nightly light curves is shown in Figure 5. Basically all nights look alike, with two prominent features: a 3-hour wave with a full amplitude in the range 0.05–0.30 mag, and rapid flickering including apparent quasi-periodic oscillations (QPOs) with *P*~1000 s.

To search for strictly periodic signals, we calculated power spectra of light curves from the discrete Fourier transform. The power spectrum of the 50-day light curve is shown in Figure 6, where the most obvious feature is the sharp spike at a frequency of 8.271(6) c/d. The peaks surrounding this signal are merely the ±1 c/d aliases, and detailed comparison of the power spectrum to the spectral window reveals no nearby periodic signals passing tests of significance.

There are also strong signals at very low frequency. In the lowest frame of Figure 7 we show an expanded view of this region, with significant peaks marked at 0.229 and 0.601 (both ±0.006) c/d. In the upper frames we show the power spectra of artificial time series, with injected signals at exactly these frequencies and sampled exactly like the actual data. Comparison of the upper frames with the peaks visible in the lowest frame shows that most of the variance in the low-frequency power spectrum arises from these signals and their aliases. We also studied time series with the aliases (rather than the correct frequencies) artificially inserted, and found that they gave much poorer matches to the real power spectrum — indicating that the true frequencies are indeed 0.229 and 0.601 c/d rather than any of the aliases.

The widths of all three power-spectrum peaks (4.37 d, 1.67 d, 0.12090 d) are consistent with the hypothesis of period constancy during the 50-day observation, but this is only provable for the latter signal.

The 8.271 c/d signal occurs at a frequency consistent with $\omega_o+N$, where $\omega_o$ is the orbital frequency [8.043 c/d (Hoard, Thorstensen, & Szkody 2000; Diaz 2001) and $N$ is the stronger low-frequency signal. The corresponding period is 2.85±0.08% shorter than $P_{orb}$ — a *negative* (or nodal) *superhump*, in the vocabulary introduced by Harvey et al. (1995).[2] $N$ is then the frequency of nodal precession. The other low-frequency signal, at 0.601 c/d, could possibly be the signature of a positive (apsidal) superhump. However, its detection is marginal, and we do not have a detection at the expected superhump frequency itself (at $\omega_o$–0.601=7.442 c/d). So that signal needs confirmation.

Figure 8 shows the synchronous summations at each frequency after "prewhitening" by subtracting best fits at the other two frequencies. The semi-amplitude of the negative superhump is 0.072(8) mag, and that of the underlying precession (4.37 d signal) is 0.083(12) mag. The putative apsidal precession (1.67 d signal) gives a complex light curve of low amplitude. Just for the record, the folded light curves in Figure 8 yield the ephemerides:

$$\begin{aligned} \text{Maximum light} &= \text{HJD } 2449857.751(3) + 0.12090(8)\ E \\ &\quad \text{HJD } 2449859.63(18) + 4.37(11)\ E \\ &\quad \text{HJD } 2449858.03(11) + 1.67(3)\ E. \end{aligned} \qquad (2)$$

However, we stress again that the detection at 1.67 d is merely a *candidate* signal.

### 3.2 DATA IN OTHER YEARS

We carried out similar studies in 1983, 2001, and 2002. Table 1 gives the main details of these campaigns. Data reduction followed the methods discussed above. These campaigns were less extensive, but the results merit a brief report. Power spectra are shown in Figure 9; these are otherwise "dirty" but have been cleaned for the aliases of the strongest signal. The strong signal of 2001 occurred at a frequency consistent with that of 1995's negative superhump. The 2002 signal occurred at 8.294(6) c/d, a slightly higher frequency; in the interpretation peddled here, this corresponds to a 10% change in the underlying nodal precession frequency. The 1983 signal occurred at 7.43±0.03 c/d. Since the latter is displaced from $\omega_o$ by –0.61±0.03 c/d, this is an apparent *apsidal* superhump and gives some support for the hypothesis that the uncertain detection of 0.60 c/d in the 1995 data is real.

## 4. PHOTOMETRY OF RX 1643+34

A similar campaign was carried out on RX 1643+34, using seven telescopes of the CBA network. This consisted of 70 nights (452 hrs) of coverage during 2002 April 10 – June 22. We used overlaps in coverage to calibrate between the various observatories, putting all this differential photometry on a common scale. The star remained in its "high state", near $V$=12.7, throughout the observations. A sample nightly light curve is shown in the upper frame of Figure 10. The obvious rapid oscillations, with an amplitude near 0.1 mag and a period near 0.01 d, are the dominant feature.  There are also dips and slow wiggles which suggest the orbital timescale, but they are not strictly periodic (even from one night to the next).

---

[2] Superhump vocabulary and zoology are briefly reviewed in Appendix A of Patterson et al. (2002), which might be consulted as a last resort. An alternative version is available at http://cba.phys.columbia.edu/zoology/.

The middle and lower frames of Figure 10 show the power spectrum of the season's light curve.  Nightly aliasing is insignificant, because of the geographical spread of observatories (Greece, Belgium, North America, Australia).  Significant detections are flagged with their frequencies in cycles/day:  a $2\omega_o$ component at 16.594(5), and signals at 0.247(8) and 8.550(6).  The latter two are respectively the nodal frequency $N$ and the negative superhump frequency $\omega_o+N$.  Inset are mean waveforms at these frequencies.  They are basically uninformative, except that the weak orbital waveform appears to resemble the "double sinusoid" associated with Roche-lobe distortion of the secondary.  All the waveforms are much noisier than expected for strictly periodic signals of this amplitude in a 13th magnitude star.  The reason is the ceaseless din of the noisy QPO, studied in Section 6 below.  The rapid and erratic variations contaminate the synchronous summations, and all the telescope aperture and data analysis in the world can't do a thing about it.

The ephemerides associated with these signals are:

$$\text{Precession maximum} = \text{HJD } 2452386.43(15) + 4.05(13) \, E$$
$$\text{Superhump maximum} = \text{HJD } 2452383.763(4) + 0.11696(7) \, E \qquad (3)$$
$$\text{Orbital minimum} = \text{HJD } 2453383.813(3) + 0.12056(2) \, E.$$

The observed semi-amplitudes are respectively 0.062(1), 0.017(3), and 0.011(2) mag.  It's very likely that only the orbital ephemeris has any real usefulness;  the other periods, associated with disk precession, are not truly stable.

We did not find any signal at or near 9.25 c/d, reported by M02. A signal of the reported amplitude would have produced a power of 350 in the lowest frame of Figure 9;  so neither this signal nor any of its aliases could have been present in 2002 at a semi-amplitude exceeding 0.007 mag.  Since the star frequently sports transient waves of similar frequency and ~0.1 mag amplitude, we suspect that this report arises from the brevity of the observation's window on that fundamentally erratic process.

## 5. SUPERHUMPS AND PRECESSION IN NOVALIKE VARIABLES

V442 Oph and RX 1643+34 are fairly good matches to the SW Sex class, characterized mainly by single-peaked emission lines, high-velocity $S$-waves, and evidence of strong outflows varying with orbital phase (Thorstensen et al. 1991, Hellier 2000, Hellier 2001).  In our study of negative superhumps, we have found a high proportion of such signals in SW Sex stars. Table 1 of Patterson (1999) lists 11 stars with negative superhumps, and 7 are confirmed or likely SWs. It may well be true that the SW Sex phenomenon is linked with nodal precession.

Some of the distinctive spectroscopic features of SWs can be reproduced by supposing that a substantial fraction of the mass-transfer stream overflows the accretion disk (Hellier & Robinson 1994, Hellier 1996).  This is a natural feature in a wobbling-disk model, since disk wobble moves the impact point above and below the disk, varying smoothly with wobble and/or orbital phase. Disk wobble also explains naturally why a signal at $\omega_o+N$ is (nearly always) accompanied by a signal at $N$: because to a distant observer, the visible disk area should change on the wobble frequency $N$.

Unfortunately, there is no simple criterion for SW Sex membership. Most novalike variables in the 3–4 hour period range (with outliers) appear to qualify. But they also tend to be "VY Sculptoris stars", a class defined more simply by the existence of fairly well-defined high and low states associated with large changes in mass transfer (Shafter 1984, Leach et al. 1999). A few novalike variables in this period range have managed to resist acquiring either label (so far). Permanent negative superhumps are actually common among all three groups, and it is still far from clear just which characteristic is causally linked to the superhumps. Since SW Sex stars have been the most extensively studied, we shall discuss that particular association in this paper. In the end, it may well turn out that all three characteristics (VY Scl, SW Sex, and permanent superhumps) are caused by a single piece of underlying physics.

## 6. PERIODIC SIGNALS AT HIGHER FREQUENCY

All light curves of both stars show prominent activity on a timescale of ~1000 s, with an amplitude varying from 0.03 to 0.3 mag. Each night's power spectrum shows "significant" features, but the frequencies rapidly change (on a timescale <<1 day). This is characteristic of a signal of very low coherence. Fourier analysis is nearly blind to these strong QPOs, since it parses the time series into components of constant frequency and amplitude.

### 6.1 V442 Ophiuchi

Nevertheless, we pressed onward. We searched for a period by averaging power spectra of 22 long nights on V442 Oph in 1995, and (separately) 25 long nights in other years. The result is seen in Figure 11. The lower frame shows a broad QPO in the range 70–105 c/d. The 1995 average in the upper frame shows a similar bump, but with stronger contributions from possibly significant features at 74.0 and 89.9 (both ±1.0) c/d. The latter are quite weak signals, just 0.01 mag semi-amplitude, much weaker than the obvious oscillations seen in the light curve.

### 6.2 RX J1643.7+3402

We carried out the same search for RX 1643+34, using the 57 nights of long (>4 hr) observation. The average nightly power spectrum is shown in Figure 12 (with the lower frame rendering the same result in log-log units). From this we learn the following:

(1) A flickering source is evident in the range 120–1000 c/d, with a continuum characterized by $P \sim \nu^{-2.2}$ ("red noise").

(2) There are features near 8 and 16 c/d as discussed above; otherwise, the distribution at low frequency is flat.

(3) In the range 70–110 c/d there is a strong and broad feature, the power-spectrum counterpart of the QPO obvious in the light curve. A narrow feature at 82.8±1.0 c/d is superimposed — a hint of a stable signal underlying the QPO. Despite its prominence in Figure 12, this is actually a very weak signal compared to the QPO. It exceeds the neighboring power by a factor of ~2, but there are ~25 independent frequency elements in the QPO; so in the original light curves, the candidate stable signal cannot generally be seen in the QPO's glare.

### 6.3 NARROWING THE PERIOD SEARCH

If truly stable signals appear, even weakly, in the nightly power spectrum, then we should be able to extract a precise period from time series over many adjacent nights. We tried this for both stars, using all data, and also with judicious selection of nights (by criteria of brightness, length and quality of observation, wavelength range). The results were basically inconclusive; the QPO noise dominated, and we could not determine a more precise period.

## 7. PRELIMINARY RESULTS ON TWO OTHER SW SEX STARS

We give here a preliminary report of periodic signals found in campaigns on two other SW Sex stars. V795 Herculis and DW Ursae Majoris are well-known members of that class (Casares et al. 1996, Shafter et al. 1988). We have been observing them sporadically over the last decade. V795 Her sported a powerful 2.8 hour photometric wave through most of the 1980s, interpreted as an apsidal superhump (Patterson & Skillman 1994, hereafter PS). This signal disappeared through 1988–1996, but by 2001 had reappeared at high amplitude. Figure 13 contains the 2002 low-frequency power spectrum, showing a powerful superhump at 8.550(6) c/d. Since the orbital frequency is 9.238 c/d (Shafter et al. 1990), this is evidently an apsidal superhump ($\omega_o-\Omega$ in superhump terminology). The weaker signals flagged in Figure 13 occur at $2\omega_o-2\Omega$, $2\omega_o-\Omega$, and $2\omega_o$.

Figure 14 shows the average nightly power spectrum in 2002, based on the 32 nights of long coverage. A broad signal is evident near 80 c/d, and it may possibly resolve into components at 71.0 and 80.4 c/d (each ±1). Similar results were obtained in our 1993 campaign, but at slightly different frequencies (PS).

The power spectra of DW UMa's seasonal light curves are shown in Figure 15. In 2002 (upper frame), the dominant signal was an apsidal superhump at $\omega_o-\Omega=6.878(6)$ c/d. The other flagged signals, in order of increasing frequency, occured at $2\omega_o-2\Omega$, $2\omega_o$, and $3\omega_o-2\Omega$. A quite different result was found in 1996 (lower frame), when the dominant signal was a nodal superhump at $\omega_o+N=7.542(3)$ c/d. The other flagged signals occurred at $\omega_o$, $2\omega_o$, $2\omega_o+N$, $3\omega_o+3N$, and two other frequencies which do not appear to bear any simple relation to $\omega_o$ (but are separated by exactly $\omega_o$).

We give these preliminary reports here to help illustrate the common (but variable on timescales of years) presence of superhumps in SW Sex stars, and the common presence of signals at higher frequency, usually in the form of a QPO. The latter signals deserve special note, since they often show components separated by exactly $\omega_o$. We now turn to this subject of kilosecond QPOs.

## 8. THE 1000 SECOND OSCILLATIONS

### 8.1  SPIN AND MAGNETISM

The presence of kilosecond QPOs is a curious phenomenon. In our many photometric campaigns on CVs, we have found them to be prominent only in confirmed or candidate SWs, and the detection rate in this class exceeds 50%. The origin of the signals remains unknown; efforts to study them are greatly hampered by the sloppiness of the QPO clock. But there are tantalizing hints in several stars that a stable signal may underlie the noise of the QPO. Available data on the QPOs is summarized in Table 4.[3]

The timescale is itself interesting. Magnetic white dwarfs in spin equilibrium should have $P_{spin}/P_{orb} \sim 0.1$ (King and Lasota 1991, Wickramasinghe, Wu, & Ferrario 1991), assuming that accretion is not primarily through a disk. Since the stars of Table 4 are all within 20% of $P_{orb}$=3.1 hrs, this identifies 1100 s as a natural value of $P_{spin}$ in such binaries — in good agreement with Table 4. Is it possible that the QPOs reflect the underlying rotation of a magnetic white dwarf?

### 8.2  COMPARISON WITH DQ HERS

Well, maybe, but then we need to understand why the main emergent signal is quasi-periodic, not strictly periodic. There are a few dozen CVs — known as the DQ Herculis stars or intermediate polars — which contain rapidly rotating magnetic white dwarfs (Patterson 1994, chapter 5.4 of Warner 1995). Their defining signature is a strictly periodic signal at $P_{spin}$. They have orbital periods similar to the SWs, and many have spin periods near 1000 s. So why should one binary join the DQ Her club, while another allies with the SW Sex stars?

The answer should be in some combination of magnetic field strength and accretion rate. The magnetospheric radius $r_{mag}$ in CVs is given by

$$r_{mag} = 2.7 \times 10^{10} \text{ cm } \mu_{33}{}^{4/7} \ \dot{M}_{16}{}^{-2/7} m_1{}^{1/7}, \qquad (4)$$

where $\mu_{33}$ is the white dwarf's magnetic moment in units of $10^{33}$ gauss×cm$^3$, $\dot{M}_{16}$ is the mass-transfer rate in units of $10^{16}$ g/s, and $m_1$ is the white dwarf mass in $M_\odot$. The separation $a$ between stars is

$$a = 3.5 \times 10^{10} \text{ cm } m^{1/3} P_{hr}{}^{2/3}, \qquad (5)$$

[3] We include here the famous "20-minute variation" of TT Ari, by far the best-known of these signals, even though the star is not yet included in lists of SW Sex stars. The reason is that spectroscopists have been controlling admission to that club, and they have been demanding phase-dependent He I absorption as the key admission credential. This seems too restrictive. In other respects TT Ari could be a member of the class, with high excitation, high and low states, strong QPOs, a hot white dwarf, positive and negative superhumps, and a typical $P_{orb}$. As noted earlier, improvements in classification may be needed before these observational terms (SW Sex, VY Scl, etc.) are reduced to physics. Anyway, we have studied TT Ari extensively and found nothing to distinguish its omnipresent QPO from those of the other entries in Table 4.

where $m$ is the total mass in $M_\odot$, and $P_{hr}$ is the orbital period in hours. Thus the condition for synchronism ($r_{mag}/a \geq 1$, an AM Her star) is essentially

$$\mu_{33} \geq 1.4 \ \dot{M}_{16}^{\ 1/2} \ P_{hr}^{\ 7/6} \qquad\qquad (6)$$

(here and below we neglect the weaker dependence on $m_1$ and $m$, since these values are usually close to 1). The great majority of AM Hers have $\dot{M}_{16} \sim 1$ and $P_{hr} \sim 1.8$, so $\mu_{33} \geq 3$ is a good rule of thumb. Polarization measurements suggest a typical $B \sim 30$ MG, and for an ordinary 0.7 $M_\odot$ white dwarf, this implies $\mu = BR^3 \approx 10^{34}$ g×cm$^3$; so this accords with the rule of thumb.

DQ Her stars have moderately well-formed disks, and disks have outer radii of $\sim 0.35a$. So the magnetospheres cannot extend out any further than $\sim 0.3a$. From (4) and (5) this implies

$$\mu_{33} \leq 0.2 \ \dot{M}_{16}^{\ 1/2} \ P_{hr}^{\ 7/6}. \qquad\qquad (7)$$

Now a few DQ Her stars have $\dot{M}$ and $P_{orb}$ comparable to the AM Hers, and in those cases (7) requires a quite low field, far below those of the AM Hers. But DQ Hers have $<P_{orb}> \approx 4$ hr, and $\dot{M}$ rises sharply with $P_{orb}$, so things get more complicated. Let us specialize to $P_{orb} = 3.5$ hr, since that is the regime of interest here. Then (7) reduces to $\mu_{33} \leq 0.8 \ \dot{M}_{16}^{\ 1/2}$.

But most DQ Hers $\dot{M}$ near $(1-3) \times 10^{17}$ g/s, suggesting $\mu_{33} \leq 3$, or lower than the AM Hers. This also agrees with the lack of strong circular polarization in DQ Hers, and the magnetic moments inferred from the assumption of spin equilibrium (see Table 2 and Section 11 of Patterson 1994). These arguments teach us that most DQ Hers are too weakly magnetic to be a plausible ancestor for AM Hers (King & Lasota 1991; Wickramasinghe, Wu, & Ferrario 1991). But several observational clues — very blue colors, weak emission, high excitation, and line profiles suggestive of strong winds — suggest that SW Sex stars may have significantly higher $\dot{M}$. With high $\dot{M}$, high values of $\mu$ can exist even for asynchronous rotators.

These ideas are summarized in Figure 16. The main parameters are $\mu$ and $\dot{M}$; for simplicity and definiteness we eliminate $P_{orb}$ by using an empirical $\dot{M}$ ($P_{orb}$) relation (Figure 7 of Patterson 1984, with $\dot{M}_{16}$ fixed at 1 below the period gap). AM Hers then must live in the upper left, above the synchronism line. DQ Hers must live below that line, but above the lower line where the magnetospheric radius is $3R_{wd}$ (since below that, a well-channeled flow to the white dwarf surface is unlikely). "Nonmagnetics" (mostly dwarf novae) live below that lower line.

The upper right region of Figure 16 is perhaps the most interesting, because stars of high $\mu$ and high $\dot{M}$ are possible ancestors of AM Hers. The latter constitute $\sim 20$–50% of all CVs, but are very strongly clustered at short $P_{orb}$ and low $\dot{M}$. A few DQ Hers — especially those with measurable polarization and relatively long spin periods — might later become AM Hers. But the ancestors should have $\mu > 10^{34}$, and this is best sought at high $\dot{M}$, where synchronism is destroyed and hard X-rays are not produced (accounting for the stars' absence in previous surveys of magnetic CVs). These could be SW Sexers. Some outer disk should still be present, to account for the superhumps; but it may be sufficiently slim, or tilted, that some of the mass-transfer flow impacts more or less directly onto the magnetosphere.

So the suggestion is that SW Sex stars contain magnetic white dwarfs drowned by a high accretion rate. If yet identified, the actual rotation periods could be the shorter of the two precise periods cited in Table 4.

### 8.3 QPOS AMONG CONFIRMED DQ HERS

If the SWs manage to make QPOs somehow from white-dwarf rotation, it would be nice if some of the known rotators did so as well. At least one does: AE Aquarii, with a 33 s rotation period known to be stable over years (Patterson 1979, de Jager et al. 1994). The signal *only manifests stability when the accretion rate is low*; at other times, when the star brightens, the stable signal is replaced by a broad QPO typically centered at ~35 s (see Figures 2 and 3 of Patterson 1979, and Figure 3 of Patterson 1994). If AE Aqr were always bright, it's a fair guess that we would see mainly the broad QPO, rather than the stable signal that underlies it. So Nature knows how to hide stable signals in the vicinity of QPOs, even if we do not yet understand it.

### 8.4 X–RAYS AND MAGNETIC CHANELLING

Magnetic white dwarfs in CVs are normally associated with strong and pulsed hard X-ray emission at $P_{spin}$. Yet SW Sex stars are very weak X-ray sources, with $F_x/F_{opt}$ typically ~0.01, among the weakest of any class of CV (Richman 1996). Can this be reconciled with the assumption of strong magnetism?

Yes, probably it can. Magnetic CVs often spill much of their accretion luminosity into the EUV/soft X-ray, partly because high density in the accreting gas enables the gas to avoid the shock above the magnetic pole and deposit accretion energy directly in the white dwarf (Kuipers & Pringle 1982, Woelk and Beuermann 1993). Frank et al. (1988) estimate the critical density to occur at $10^{-7}$ g/cm$^3$, and King & Lasota (1991) calculate that accreting gas at the surface must have a minimum density given by

$$\rho_{min} = 4.6 \times 10^{-10} \, \mu_{33}^{2} \, P_3^{-5/3} \text{ g/cm}^3, \qquad (8)$$

where $P_3$ is the spin period in units of $10^3$ s. Thus for a 1000 s spin period, the transition to EUV/soft X-ray occurs at $\mu_{33}$=10, or $B$=30 MG for an ordinary 0.7 $M_\odot$ white dwarf. This is a plausible dipolar field strength, indeed a typical field for an AM Her binary.

However, this cannot be the whole story, because strictly periodic signals are not seen at *any* wavelength. A more promising "pulsar" model could release much of the accretion energy at the outer edge of the magnetosphere, where a natural clock is provided by white-dwarf spin but heavily buffeted by the orbital timescales in (and response timescales in) the accreting gas. Some of this gas is likely to be ejected, but some should be accreted too, in order to provide the long-term heating required by the high white-dwarf temeperatures observed in quiescent SWs.

The spacing of high-frequency components by exactly $\omega_o$ and $2\omega_o$, observed in at least 4 SWs, is another tantalizing clue. This first made us think of the DQ Hers, where this spacing is very common due to the reprocessing of pulsed X-rays/UV in the secondary (and other structures fixed in the orbital frame). Does this structure indicate a stable underlying clock? No, not necessarily. More accurately, it suggests that the secondary sees a fairly strong incident flux pulsed (in fact, sweeping prograde) at the same period that we observe — and then reprocesses it at the lower sideband. Several variants on this can easily occur from the specific geometry, discussed for DQ Hers by Warner (1986). This pulsed signal must be strong, or its daughter signal at the lower sideband would be invisible; but it's unlikely to be coherent, since the signal *we* observe is not. It should, however, be just as coherent as the parent signal, since the reprocessor follows a strict clock (the orbit).

## 9. OTHER SIGNS OF MAGNETISM, REAL AND IMAGINED

So far, the only real evidence cited here for magnetism is the kilosecond QPOs. While these appear to be characteristic of SWs, and are merrily clustered near 0.1 $P_{orb}$, they still do not amount to *compelling* evidence since the timescales are rough and quite possibly characteristic of the disk (*sans* magnetism) as well. What other evidence might exist?

There are two *ne plus ultra* indicators of magnetically channeled accretion. One is periodic circular polarization, the usual credential for AM Her classification. And indeed, this has been recently reported for two SW Sex stars (Rodriguez-Gil et al. 2001a, 2001b). But the observed periodic term is quite weak (~0.2%) and the data are not yet sufficiently extensive to establish stability, even over a few nights. (In other words, this could be yet another manifestation of the QPO.) It is possible that future observations of this type will clinch the matter, but the needed evidence is not yet in.

The other indicator is a period of demonstrably high phase stability in the photometry (at any wavelength, but X-ray and optical are the most common and likely). This is the usual credential for DQ Her classification. Over hundreds of nights, we have tried very hard to find such a period in SWs. We have even succeeded in identifying candidate signals consistent with strict phase stability over ~30 nights (AH Men, V795 Her, DW UMa). However, the resultant signal is so weak that when we split it into a dozen independent segments, the QPO noise dominates and the evidence for phase stability disappears or becomes ambiguous. So this effort is still not conclusive. To achieve success, we need

(1) to observe one of these stars even more assiduously;

(2) to find a wavelength less contaminated by the QPO; or

(3) to get lucky and find one of these stars in a state of low (but detectable) accretion and low QPO acitivity.

Since our main research instrument is a network of very small telescopes, only the first is typically feasible for us; but we recommend all three approaches to everyone!

## 10. NEGATIVE SUPERHUMPS AND DISK WARP

To explain negative superhumps with regressing nodes, we should understand what makes the disk non-coplanar with the orbit. The most well-studied process for warping the disk invokes the effect of radiation on the inner disk (Iping & Petterson 1990, Pringle 1996, Maloney et al. 1998). This works well for luminous X-ray binaries but requires very strong radiation from the central object, and hence is probably not relevant for these white dwarfs of modest luminosity. Also possible is a dynamical instability at the 3:1 resonance in the disk. Lubow (1992) discovered and studied this instability, but concluded that its weakness and slow growth rate made it unlikely to be important; Murray and Armitage (1998) reached a similar conclusion.

Another possibility (of course) is magnetism. An inclined magnetic dipole centered on the white dwarf breaks the azimuthal asymmetry and provides a path vertically out of the disk plane. Detailed studies of this geometry have been presented in the context of X-ray binaries (Lai 1999) and T Tauri stars (Terquem & Papaloizou 2000); these have shown that an inner-disk warp can be created and lead to a global retrograde precession. Murray et al. (2002) alternatively invoke magnetism of the secondary and obtain the same result, with a large warp situated in the outer disk (see their Figure 4 for a vivid illustration).

Each of these theories has a specific observational point in its favor:

(1) All the observed negative superhumps are "permanent", i.e. endure for months to years. None grow or decay rapidly, as the positive superhumps (famous in dwarf novae) are wont to do. Thus a slow growth rate is not necessarily disqualifying.

(2) White-dwarf magnetism is demonstrably present through X-ray pulses in two negative superhumpers (TV Columbae and V709 Cassiopeiae), and is arguable on the basis of QPO evidence in most of the others (catalogued here and in Table 1 of Patterson 1999).

(3) Secondary-star magnetism is always plausible in CVs, since these are cool, rapidly rotating stars. The magnetism in such stars could well have built-in 5–20 year timescales, which might explain the slow transitions between negative and positive superhumps.

In view of the subtleties raised by these points, however, and in view of the need to steer our other results through to proper publication, we now quickly escape, and leave these issues as fodder for intrepid theorists!

## 11. SUMMARY AND OUTLOOK

1. We report spectroscopy of RX 1643+34, showing a blue continuum, weak emission lines of high excitation, high-velocity components in the Balmer lines, and phase-dependent He I absorption. The orbital period is 0.120560(14) d. These basically establish its membership in the SW Sex class of cataclysmic variables.

2. We report long photometric campaigns on RX 1643+34 and V442 Oph, another SW Sex star.

Both stars show negative superhumps, with periods respectively 3.2±0.2% and 2.9±0.2% shorter than $P_{orb}$. A positive superhump is also possible in V442 Oph, with a period 8.2±0.6% longer than $P_{orb}$.

3. Both stars show photometric signals at the beat frequency between orbit and superhump. This follows the usual pattern seen in negative superhumpers (N, $\omega_o$, $\omega_o+N$). V442 Oph may also show the beat frequency between orbit and positive superhump (i.e., a signal at $\Omega$).

4. We also briefly report long photometric campaigns on two other SW Sex stars, V795 Her and DW UMa. The former showed a return of its positive superhump, unseen since the 1980s, and the latter showed powerful superhumps which changed sign (from negative to positive) between 1996 and 2002.

5. These superhump detections follow the patterns familiar from studies of dwarf novae (e.g. Figure 1 of Patterson 1999). Thus they furnish good evidence that *SW Sex stars have disks*. The point is of some importance, since other evidence is somewhat ambiguous (single-peaked and uneclipsed emission lines, peculiar eclipse waveform).

6. It's possible that the signal at $N$ arises directly from the varying area of a wobbling disk. Assuming that retrograde nodal precession is the underlying phenomenon, the signal at $\omega_o+N$ occurs at the frequency with which disk-secondary geometry recurs. Thus the origin of the superhump could be as simple as heating of the secondary by radiation from the disk (which reaches the secondary relatively unimpeded, since the disk is not coplanar).

7. We see long-term variability in the superhump behavior of essentially all the SWs, and candidates, we observe over a baseline ≥ 6 years. The stars tend to linger for years at a time in states of positive, negative, and no superhumps. Since superhumps and line profiles both reflect deep properties of the disk (maybe even the *same* deep property), it would be fascinating to know if the spectroscopic signatures followed suit. Repeated spectroscopic studies in different superhump states would be of high interest.

8. Three of the four stars show strong kilosecond QPOs, and three of the four show a two-peak structure with components separated by $\omega_o$ and $2\omega_o$. Other examples are given in Table 4. These could represent DQ Her-type rotation periods, with the lower-frequency signal attributed to reprocessing of pulsed flux in structures fixed in the orbital frame. This raises two questions: why are the emergent signals quasiperiodic rather then periodic, and why is the high-energy pulsed flux hidden from direct observation? We discuss these issues without resolution. But among magnetic CVs, this is a plausible outcome when high $\dot{M}$ is coupled with high $B$. The SW Sex stars could be ancestors of AM Her stars, spun up to kilosecond periods and drowned by a high accretion rate. If true, this would fill an important gap in the understanding of CV evolution.

9. Of really critical interest would be *observations in different luminosity states*, because that provides a good chance to be rid of the vexing QPO.

10. So there are some whispers of white-dwarf magnetism: QPOs near 0.1 $P_{orb}$, some evidence

for underlying stable signals, frequency spacing by $\omega_o$ and $2\omega_o$, polarimetry, negative superhumps suggesting disk wobble, and the need for a still unidentified class of high-$B$ stars to provide an ancestor channel for AM Her stars. These appear now to warrant a few pages in a scientific journal. But the really important work lies ahead: compelling evidence for magnetism, and a theory to weave all these interesting threads into a coherent whole!

We thank Jim Hannon, Dieter Husar, Alina Hirschmann, Lucas Cieza, Donn Starkey, and Jonathan Kemp for contributions of data to this enterprise. The work was supported in part by grants 00-98254 and 99-87334 to Columbia University and Dartmouth College, and Research Corporation grant GG-0042 to the Center for Backyard Astrophysics.

TABLE 1
RX J1643.7+3402 Emission Features − 2002 June

| Feature | E.W.[a] (Å) | Flux[b] ($10^{-15}$ erg s$^{-1}$ cm$^{-2}$ Å$^{-1}$) | FWHM[c] (Å) |
|---|---|---|---|
| H$\gamma$ | 2.4 | 205 | 13 |
| $\lambda$4648 | 2 | 105 | 24 |
| HeI $\lambda$4686 | 2.0 | 22 | 21 |
| H$\beta$ | 4.1 | 228 | 15 |
| HeI $\lambda$4912 | 0.5 | 26 | 13 |
| HeI $\lambda$5015 | 0.4 | 19 | 11 |
| H$\alpha$ | 16.1 | 336 | 21 |
| HeI $\lambda$6678 | 1.2 | 25 | 15 |
| HeI $\lambda$7067 | 0.7 | 12 | 16 |

[a] Emission equivalent widths are counted as positive.
[b] Absolute line fluxes are uncertain by a factor of about 2, but relative fluxes of strong lines are estimated accurate to ~10 per cent.
[c] From Gaussian fits.

TABLE 2
Fits to Radial Velocities[a]

| Data Set | $T_0$[b] | $P$ (d) | $K$ (km s$^{-1}$) | $\gamma$ (km s$^{-1}$) | $N$ | $\sigma$ (km s$^{-1}$) |
|---|---|---|---|---|---|---|
| Hβ emisssion | 407.945(2) | 0.120560(14) | 126(13) | −68(10) | 141 | 57 |

[a] Fits are of the form $\upsilon(t) = \gamma + K \sin [2\pi (t - T_0) / P]$. The number of points used in $N$ and $\sigma$ is the standard deviation from the best fit.

[b] Blue-to-red crossing, HJD − 2,452,000.

TABLE 3
LOG OF PHOTOMETRY

| Telescope | Observer | Nights/hours (years) | Telescope | Observer | Nights/hours (years) |
|-----------|----------|----------------------|-----------|----------|----------------------|
| A. V442 Ophiuchi | | | B. RX J1643.7+3402 (2002) | | |
| SAAO 0.75 m | D. O'Donoghue | 5/17 (1983) | CBA–Flagstaff 41 cm | R. Fried | 21/127 |
| CTIO 1.0 m | J. Patterson, J. Kemp | 18/99 (1995) | CBA–East 66 cm | D. Skillman | 13/72 |
| CBA–West 35 cm | D. Harvey | 10/38 (1995) | CBA–Colorado 35 cm | E. Beshore | 7/43 |
| CBA–East 66 cm | D. Skillman | 7/22 (1995) | CBA–Belgium 35 cm | T. Vanmunster | 12/54 |
| CBA–Nelson 35 cm | R. Rea | 14/54 (2001–2) | CBA–Townsville 20 cm | N. Butterworth | 4/22 |
| CBA–Utah 51 cm | J. Foote | 8/37 (2002) | NOA/KAS 76 cm | P. Niarchos | 6/40 |
| CBA–Flagstaff 41 cm | R. Fried | 3/15 (2002) | CBA–Concord 44 cm | L. Cook | 3/12 |
| CBA–Townsville 20 cm | N .Butterworth | 2/8 (2002) | CBA–Indiana 35 cm | D. Starkey | 2/10 |
| CBA–Perth 35 cm | G. Bolt | 3/13 (2002) | CBA–Huntley 28 cm | O. Brettman | 1/4 |
| CBA–Pretoria 31 cm | B. Monard | 12/76 (2002) | CBA–Connecticut 20 cm | J. Hannon | 1/3 |
| C. V795 Herculis (2001, 2002) | | | D. DW UMa (1996, 2002) | | |
| CBA–East 66 cm | D. Skillman | 20/100 | CBA–East 66 cm | D. Skillman | 22/139 |
| CBA–Belgium 35 cm | T. Vanmunster | 14/47 | CBA–Utah 51 cm | J. Foote | 4/32 |
| CBA–Flagstaff 41 cm | R. Fried | 3/21 | CBA–Belgium 35 cm | T. Vanmunster | 4/31 |
| CBA–Concord 44 cm | L. Cook | 10/40 | KUC 32 cm | B. Martin | 2/14 |
| CBA–Utah 51 cm | J. Foote | 12/72 | NOA/KAS 76 cm | P. Niarchos | 2/10 |
| CBA–Huntley 28 cm | O. Brettman | 2/7 | SARA 90 cm | M. Wood | 2/16 |
| CBA–Hamburg 41 cm | D. Husar | 1/4 | CBA–Flagstaff 41 cm | R. Fried | 2/14 |
| CBA–Indiana 35 cm | D. Starkey | 1/3 | | | |

TABLE 4
Kilosecond QPO Data

| Star | QPO Period (s) | Stable Components?[a] (s) | References |
|---|---|---|---|
| V442 Oph | 1000 | 960/1170 | This paper. |
| RX 1643+34 | 1000 | 1040 | This paper. |
| V795 Her | 1150 | 1160/1310 | PS, this paper, Rodriguez-Gil et al. 2001b. |
| DW UMa | | 2974/2375 | This paper. |
| WX Ari | 1180 | | This paper[b]. |
| AH Men | 1100 | 1040/1270 | Patterson 1995. |
| TT Ari | 1200 | 1168 | Smak & Stepien 1969, Kraicheva et al. 1999b, this paper[b]. |
| V751 Cyg | 1230 | | Patterson et al. 2001. |
| LS Peg | 1240 | 1800 | Taylor et al, 1999, Rodriguez-Gil et al. 2001a. |
| BH Lyn | (1030) | | This paper[b]. |
| MV Lyr | (2800) | | Borisov 1992, Skillman et al. 1995, Kraicheva et al. 1999a. |
| DW Cnc | (4400) | | Uemura et al. 2002. |

[a] Strict stability in phase has not been demonstrated for any of these signals. Some are presumably random, transient features in the powerful QPO. Probably *most* of them are.

[b] These refer to unpublished data being prepared for publication.

## FIGURE CAPTIONS

FIGURE 1. — Mean flux-calibrated spectrum of RX J1643.7+3402 from 2002 June.

FIGURE 2. — Period search of Hβ velocities of RX J1643.7+3402. The full periodogram shows fine-scale ringing from the gap between 2002 May and June runs; this figure was constructed by connecting local maxima in the periodogram with straight lines.

FIGURE 3. — Hβ velocities of RX J1643.7+3402 folded on the adopted ephemeris, the zero point o which is the shallower of the two minima in the orbital light curve. Round dots are from 2002 June, and small squares from 2002 May. The uncertainties shown are estimated from counting statistics. Date are repeated once for continuity.

FIGURE 4. — Spectra of RX J1643.7+3402 from 2002 May (top) and 2002 June (bottom), rectified, phase-averaged, and displayed as greyscale images. Images are negatives (black=bright). The phase increases vertically, and the data are repeated once for continuity.

FIGURE 5. — Light curves of V442 Oph in 1995 at $V$=14, its "high state". Essentially all nights look alike, with rapid flickering, 1000 s QPOs, and a 3-hour wave of variable amplitude.

FIGURE 6. — Power spectrum of the 50-night V442 Oph light curve, with strong signals marked with their frequencies in cycles/day (±0.006).

FIGURE 7. — *Lower frame:* low-frequency 1995 power spectrum of V442 Oph (magnified view of Figure 2). *Middle frame:* power spectrum of an artificial time series with a 0.229 c/d signal inserted, demonstrating that the other strong peaks in the lower frame (0.77, 1.24) are merely aliases of the strong signal. *Upper frame:* power spectrum of an artificial time series with 0.601 c/d, showing weaker features present in the real power spectrum of the lower frame (and therefore tending to support the independent existence of a signal at 0.601).

FIGURE 8. — Mean light curves of V442 Oph at the three candidate periods. The folding ephemerides are given in Eq. (2).

FIGURE 9. — Other seasonal power spectra of V442 Oph, cleaned for the aliases of the strongest signal. The negative superhump dominated in 1995, 2001, and 2002; but the 1983 result suggests a possible apsidal superhump.

FIGURE 10. — The 2002 photometry campaign on RX 1643+34. *Upper frame:* a sample nightly light curve, dominated by obvious rapid variations with P~0.01 d. *Middle and lower frames:* power spectrum of the 35-night light curve, where periodic signals are marked with their frequency in cycles/day. Daily aliases are marked with "A". Inset figures show mean waveforms at these frequencies (nodal precession, superhump, and orbit).

FIGURE 11. — Average nightly power spectra of V442 Oph. *Lower frame,* average of 25 long nights (>4.5 hours) in 1983, 2001, and 2002, showing a broad QPO in the range 70–105 c/d. *Upper frame,* average of 22 long nights in 1995 (these runs are of better quality since they were taken in excellent conditions on a large telescope). Possibly significant narrow features are flagged with their frequency, and have a spacing consistent with $2\omega_o$.

FIGURE 12. — Average nightly power spectrum of RX 1643+34. Visible features in the upper frame are superhump/orbital signals near 8 and 16 c/d, a strong QPO centered on 85±3 c/d, and a possible sharp feature near 82.6 c/d. The marked features have an error of ±1 c/d. The lower frame is a log–log plot of the same data.

FIGURE 13. — Power spectrum of the V795 Her light curve in 2002, "dirty" except for cleaning of the aliases of the strong signals. The latter are flagged with their frequencies in c/d (±0.006). The main superhump at 8.550 c/d rises off-scale to a power of 800, and its mean waveform is inset.

FIGURE 14. — *Upper frame,* the nightly power spectrum of V795 Her in 2002, averaged over the 32 nights of long coverage (>5 hours). A bump signifies a QPO near 80 c/d, and it may resolve into components at 71.0 and 80.4 c/d (each ±1.0). *Lower frame,* the same power spectrum in logarithmic units. The large bumps at low frequency arise from the powerful signals near 8.5 and 17 c/d (evident in Figure 13).

FIGURE 15. — Seasonal power spectra of DW UMa in its high state. *Upper frame:* the 23-night result in 2002, "dirty" except for the cleaning of aliases of strong signals. A powerful (apsidal) superhump dominates at $\omega_o - \Omega = 6.878 \pm 0.006$ c/d. Other flagged signals occur, in order of increasing frequency, at $2\omega_o - 2\Omega$, $2\omega_o$, and $3\omega_o - 2\Omega$. *Lower frame:* the 41-night result in 1996, "dirty" except for the cleaning of aliases of the strong signals. A powerful (nodal) superhump dominates at $\omega_o + N = 7.542 \pm 0.003$ c/d, and rises off-scale to a power of 300. Inset is its mean waveform. The flagged signals occur, in order of increasing frequency, at $\omega_o$, $\omega_o + N$, $2\omega_o$, $2\omega_o + N$, $3\omega_o + 3N$, and two unidentified frequencies separated by exactly $\omega_o$.

FIGURE 16. — Approximate location of magnetic CVs in $\mu - \dot{M}$ space. The weak dependence on $m$ and $m_1$ is ignored (=1), and $P_{orb}$ is eliminated through an empircal $\dot{M}$ ($P_{orb}$) relation. See text for discussion. AM Her stars live above the upper line, and nonmagnetic CVs live below the lower line. DQ Hers and SW Sexers live in between; see text for discussion.

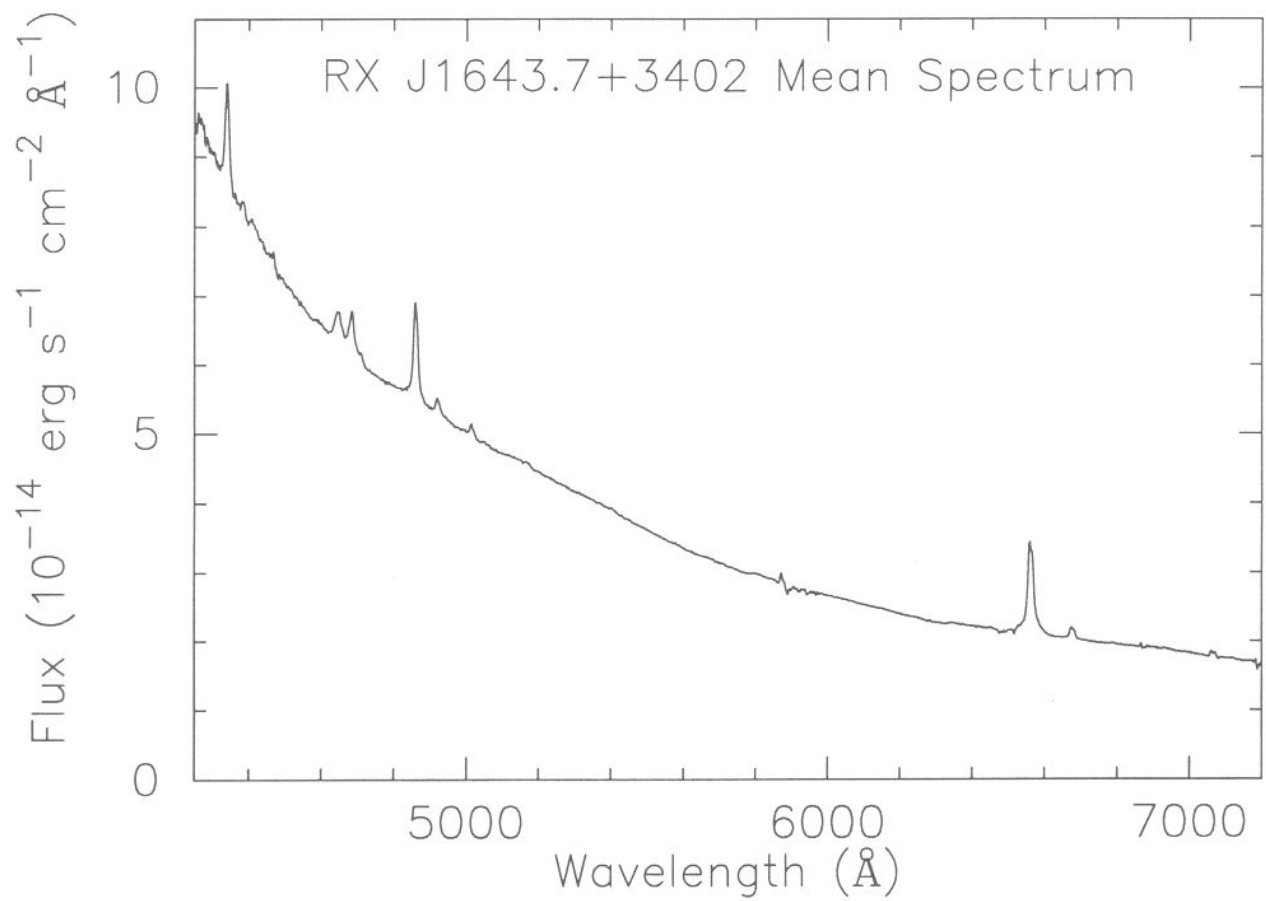

Fig. 1.— Mean flux-calibrated spectrum from 2002 June.

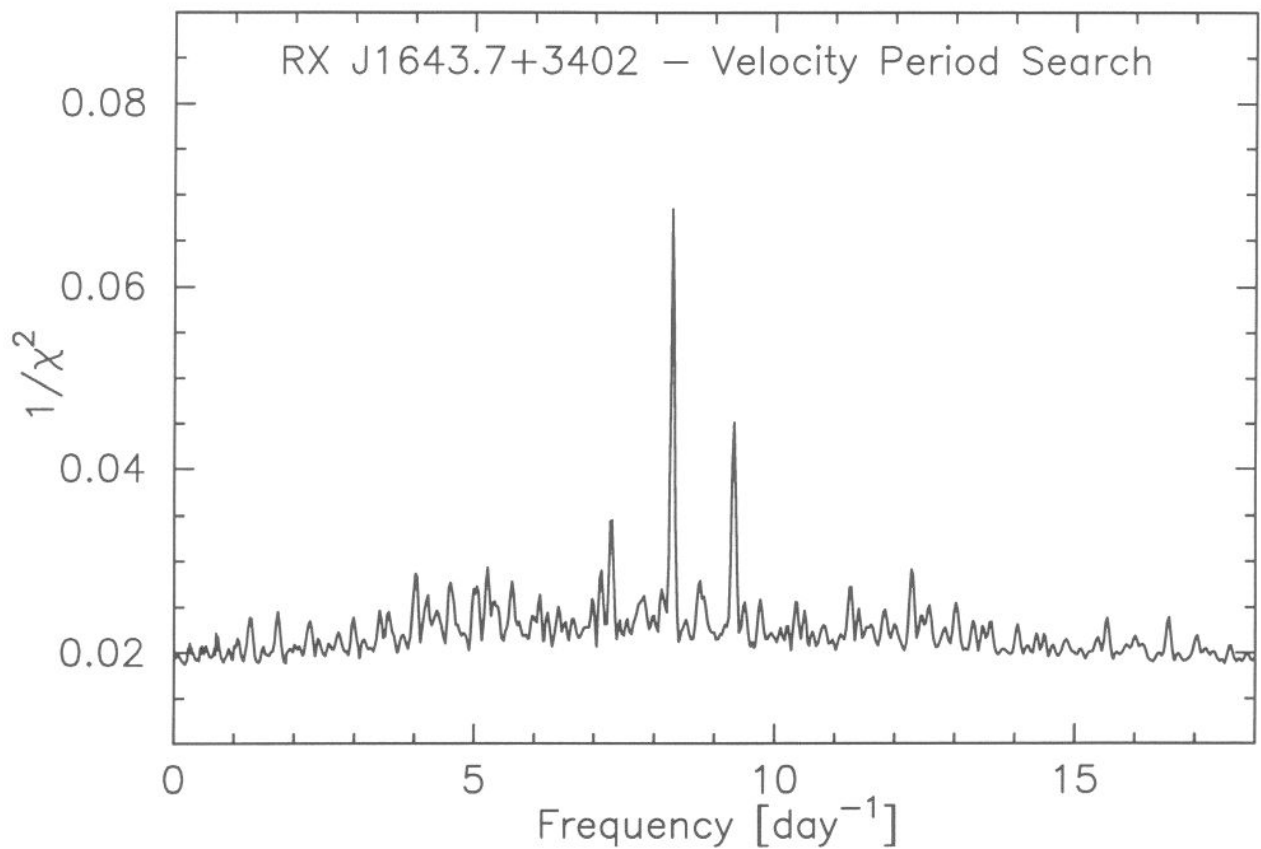

Fig. 2.— Period search of Hβ velocities of RX J1643.7+3402. The full periodogram shows fine-scale ringing from the gap between the 2002 May and June runs; this figure was constructed by connecting local maxima in the periodogram with straight lines.

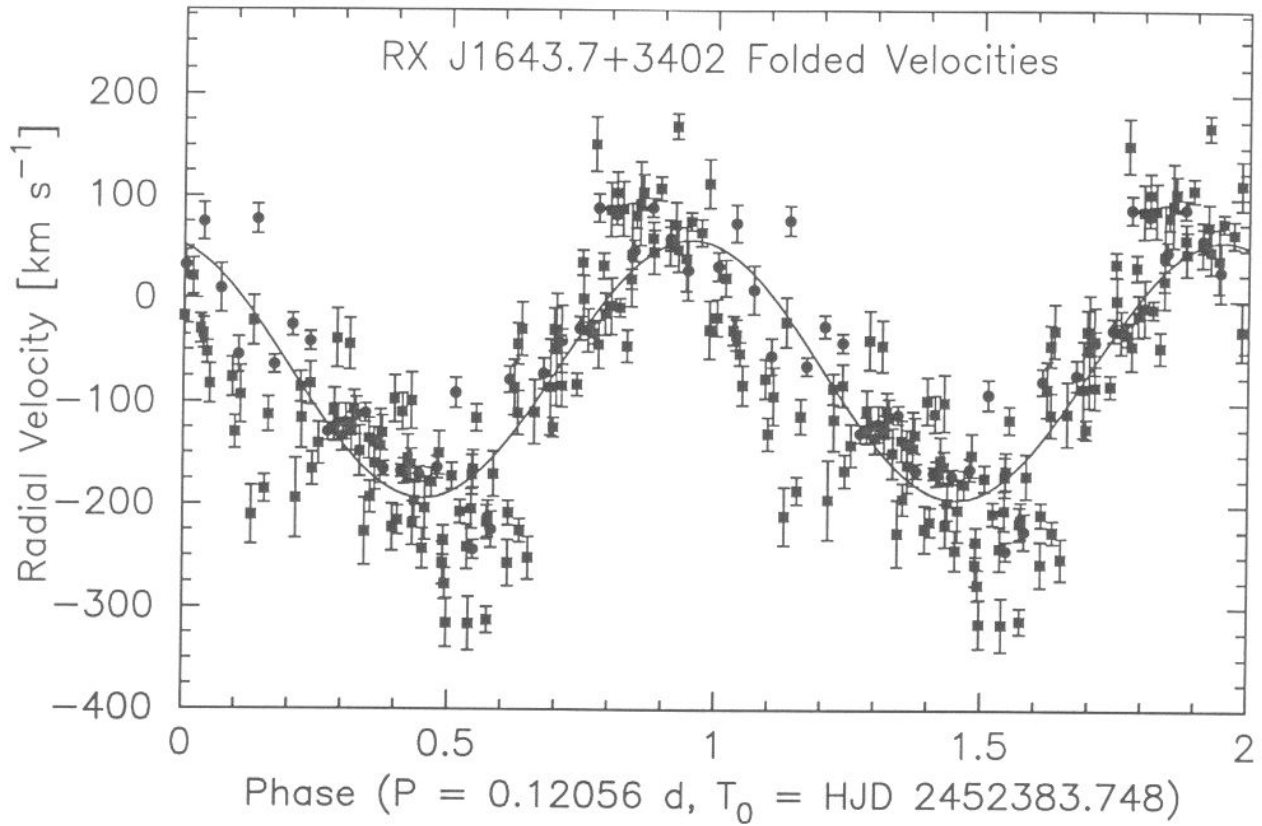

Fig. 3.— Hβ velocities of RX J1643.7+3402 folded on the adopted ephemeris, the zero point of which is the shallower of the two minima in the orbital light curve. Round dots are from 2002 June, and small squares from 2002 May. The uncertainties shown are estimated from counting statistics. Data are repeated once for continuity.

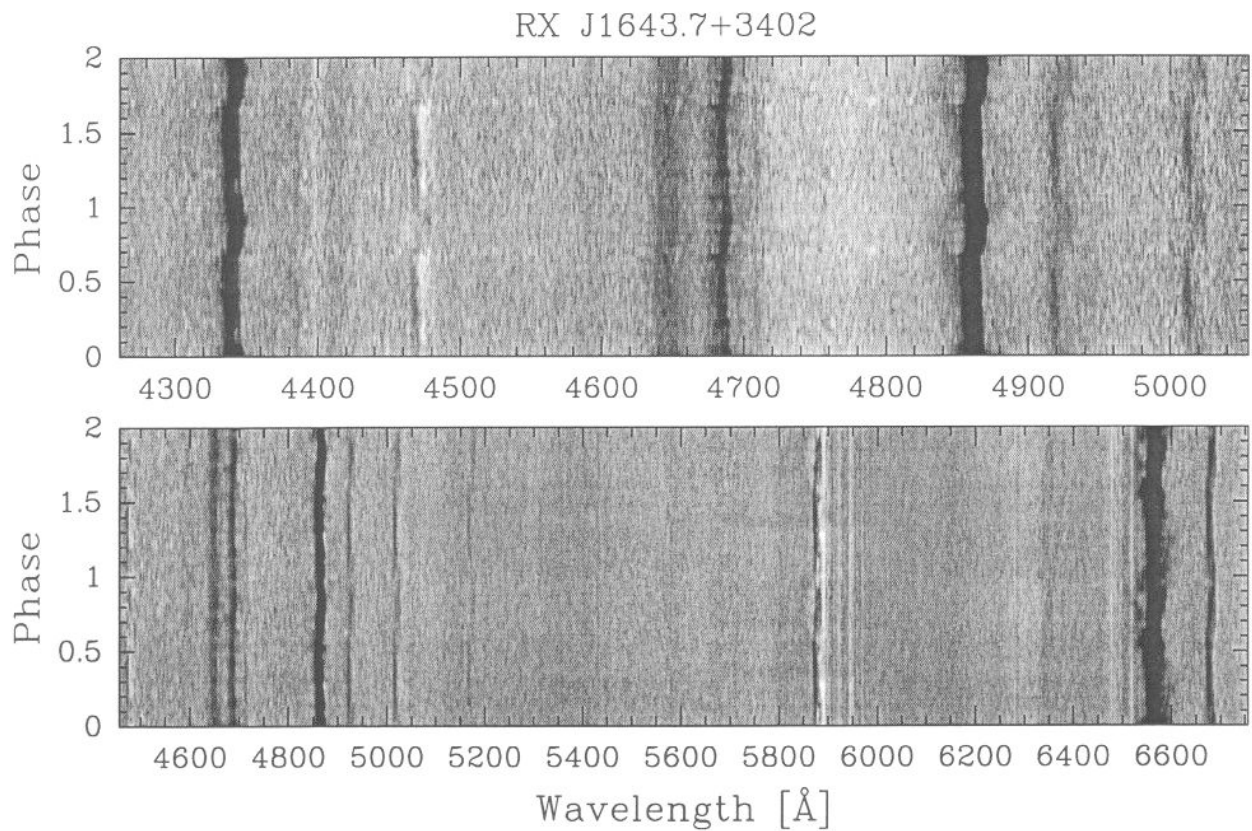

Fig. 4.— Spectra from 2002 May (top) and 2002 June (bottom), rectified, phase-averaged, and displayed as greyscale images. Images are negatives (black = bright). The phase increases vertically, and the data are repeated once for continuity.

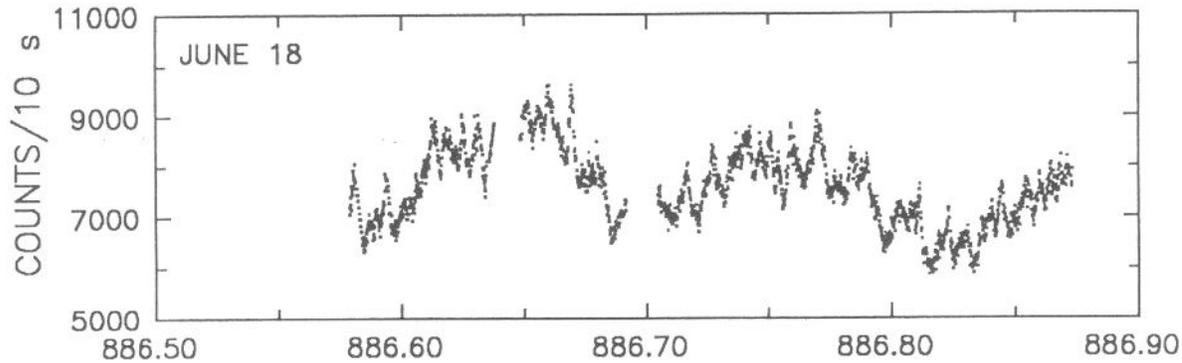

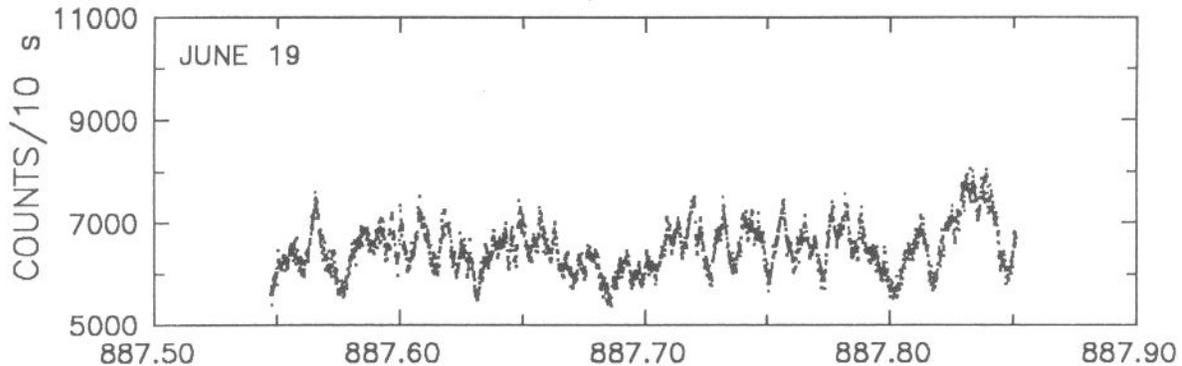

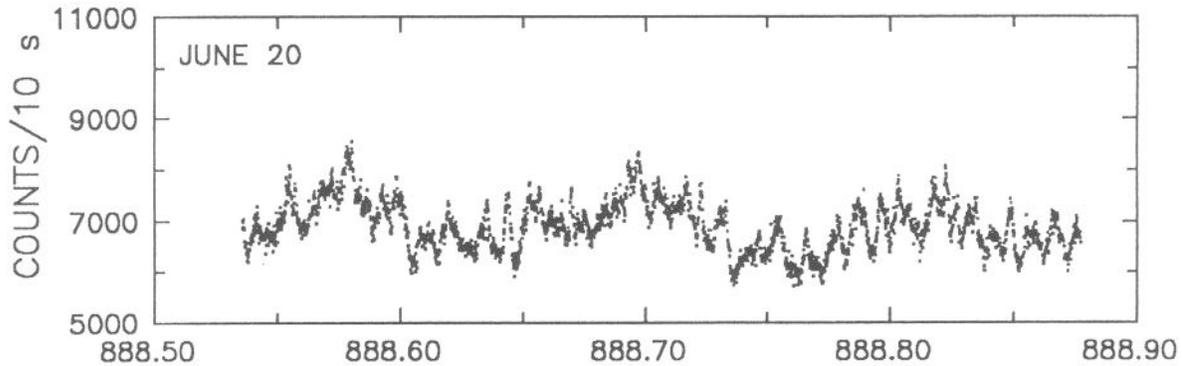

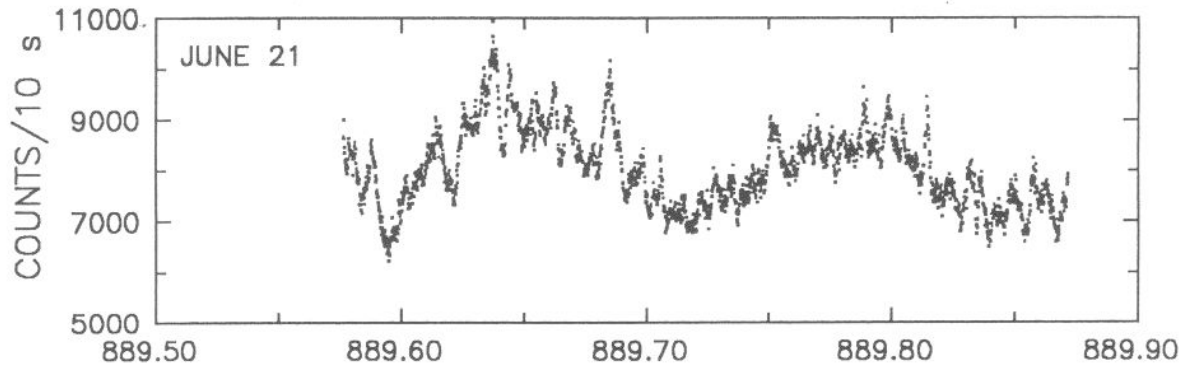

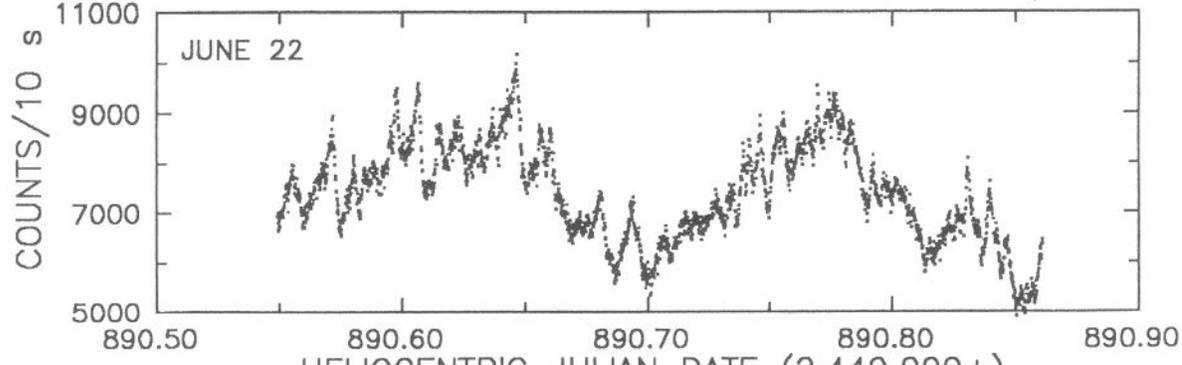

HELIOCENTRIC JULIAN DATE (2,449,000+)

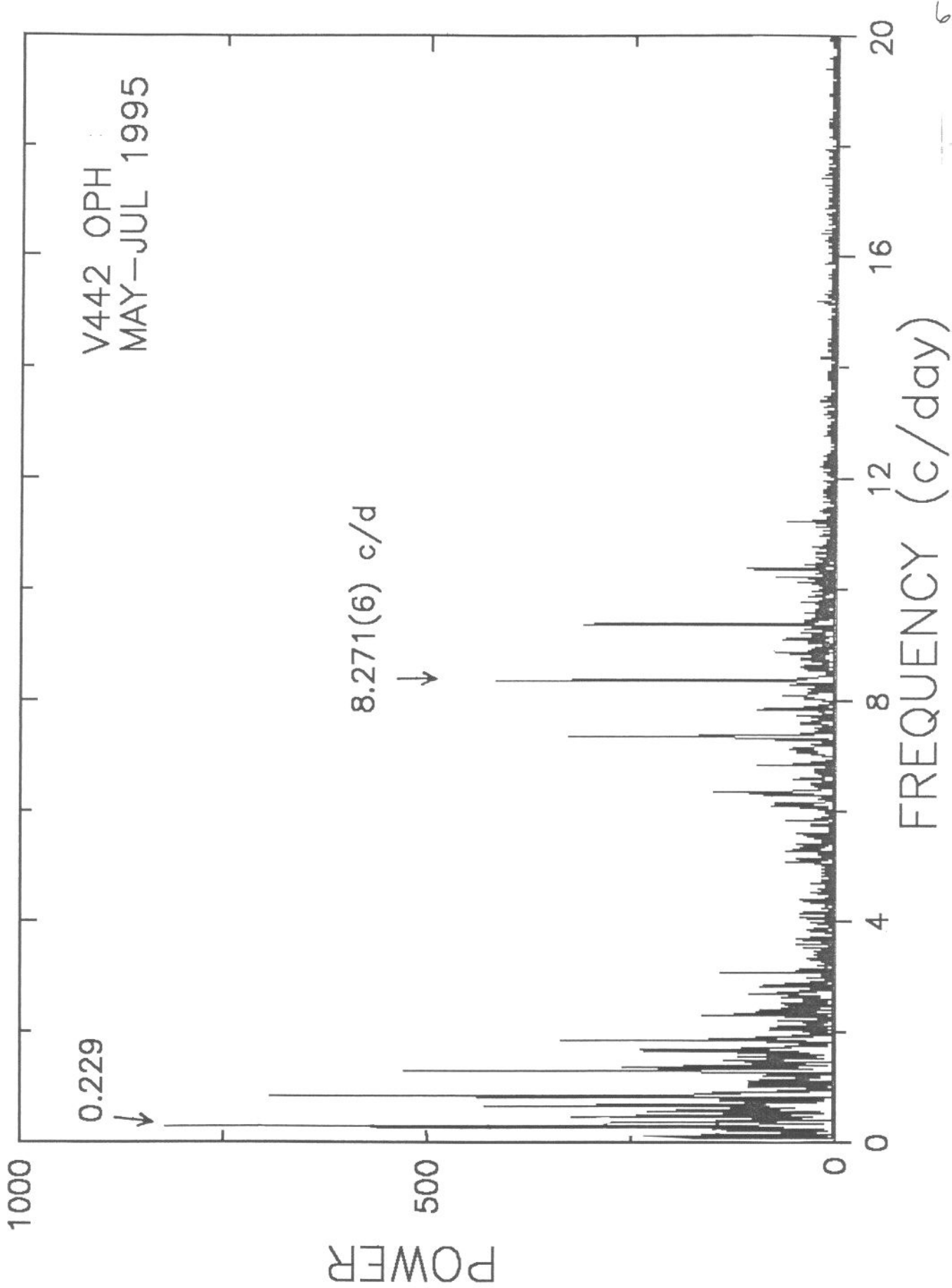

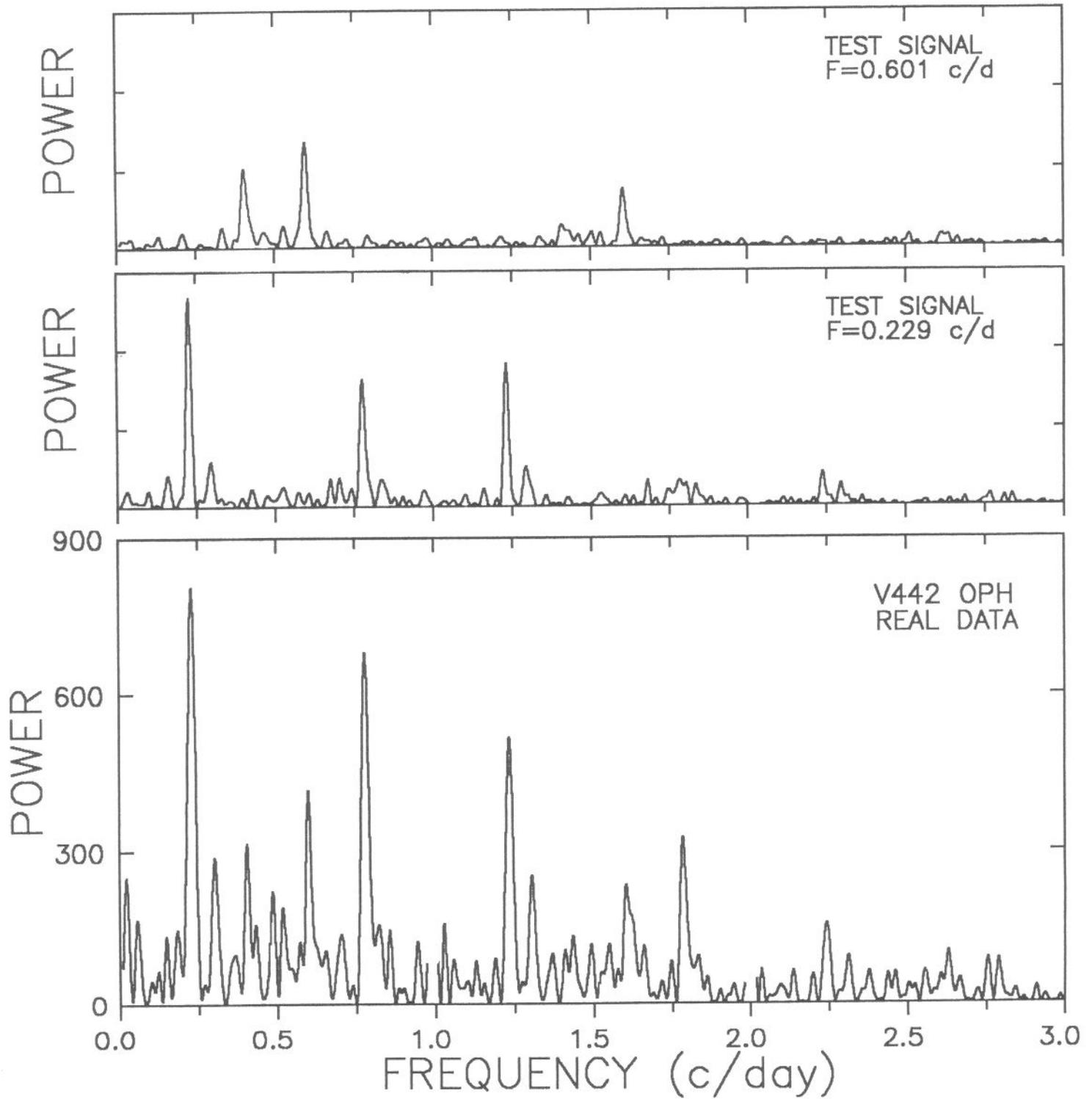

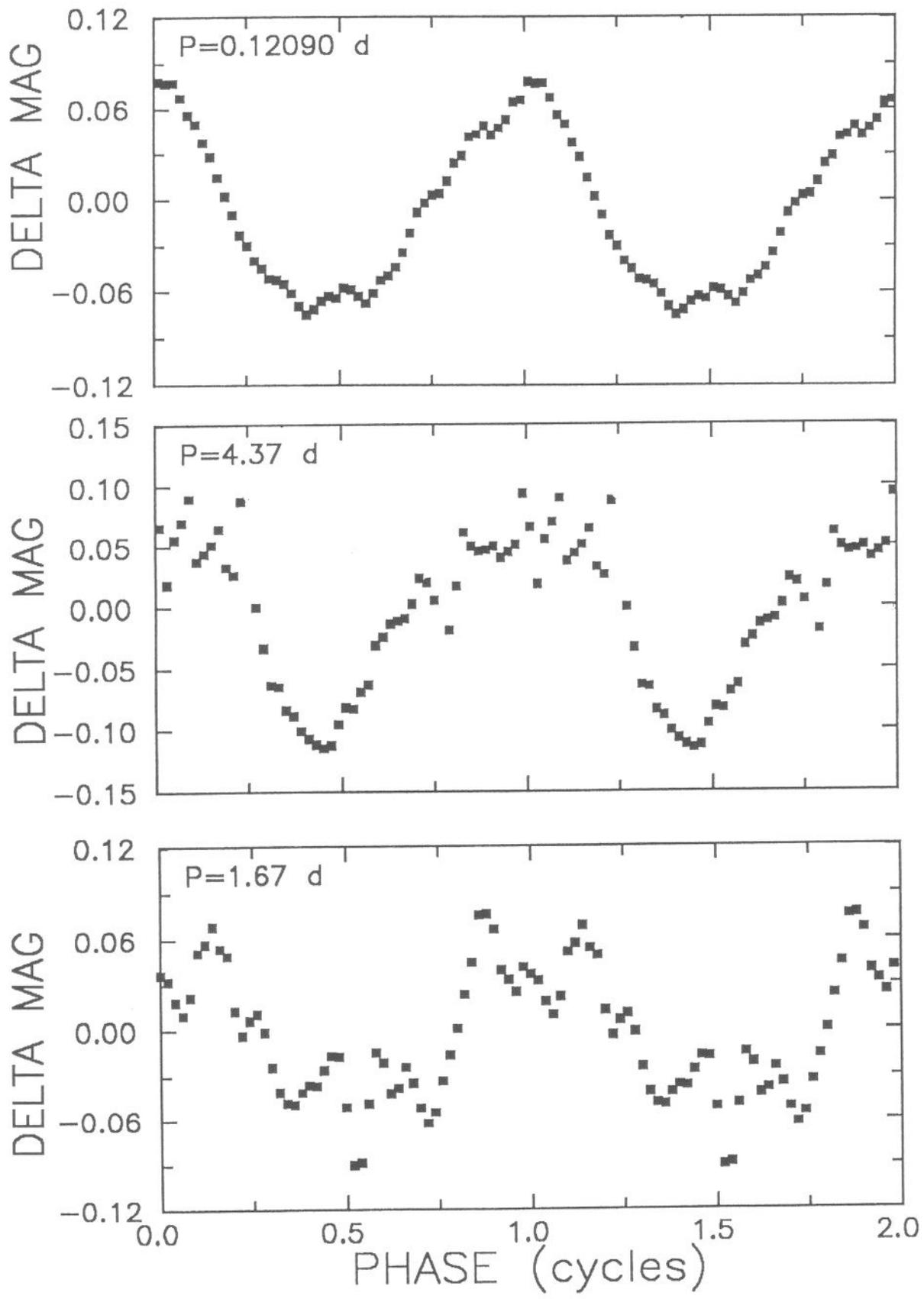

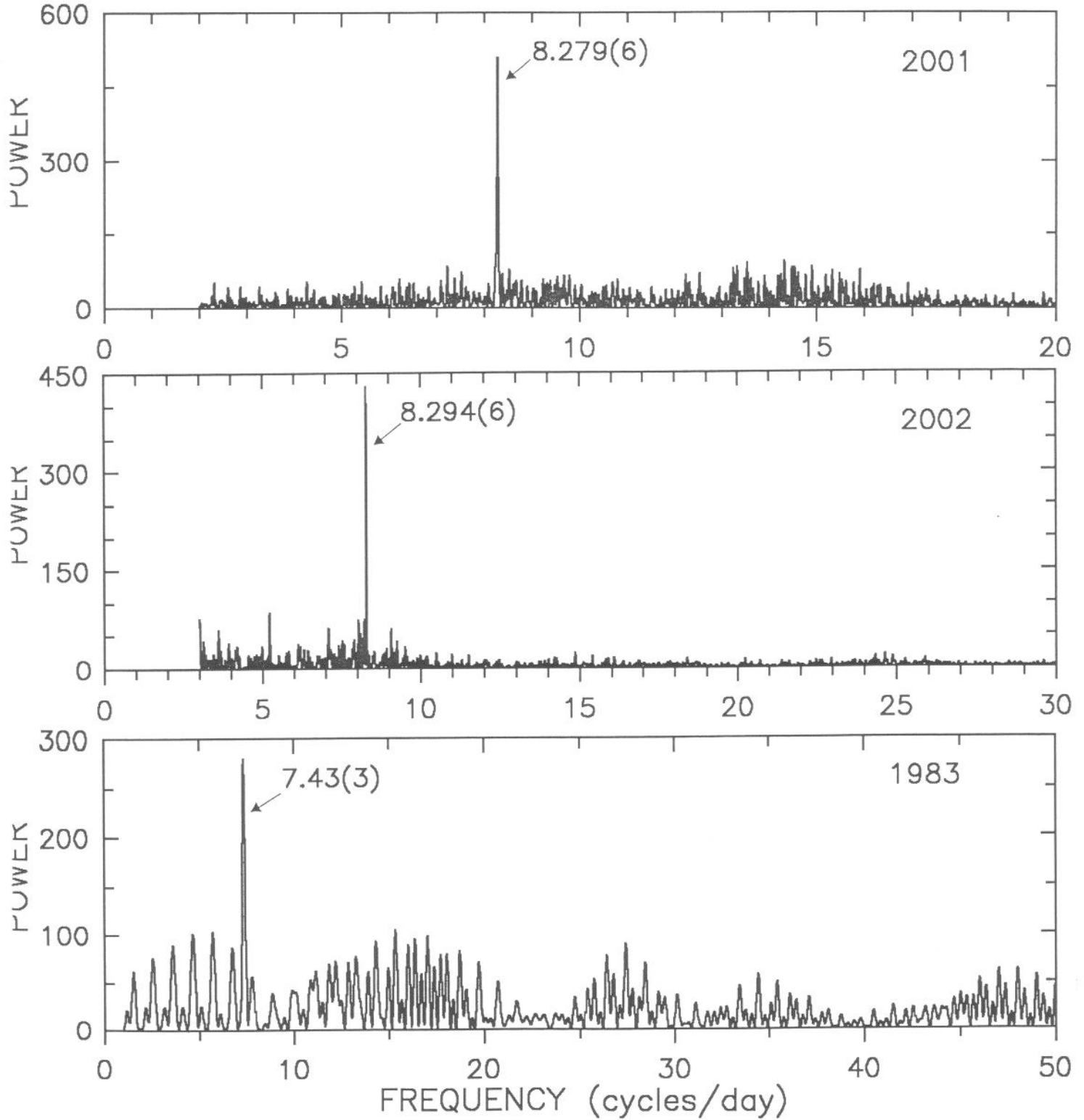

OTHER SEASONAL POWER SPECTRA OF V442 OPH

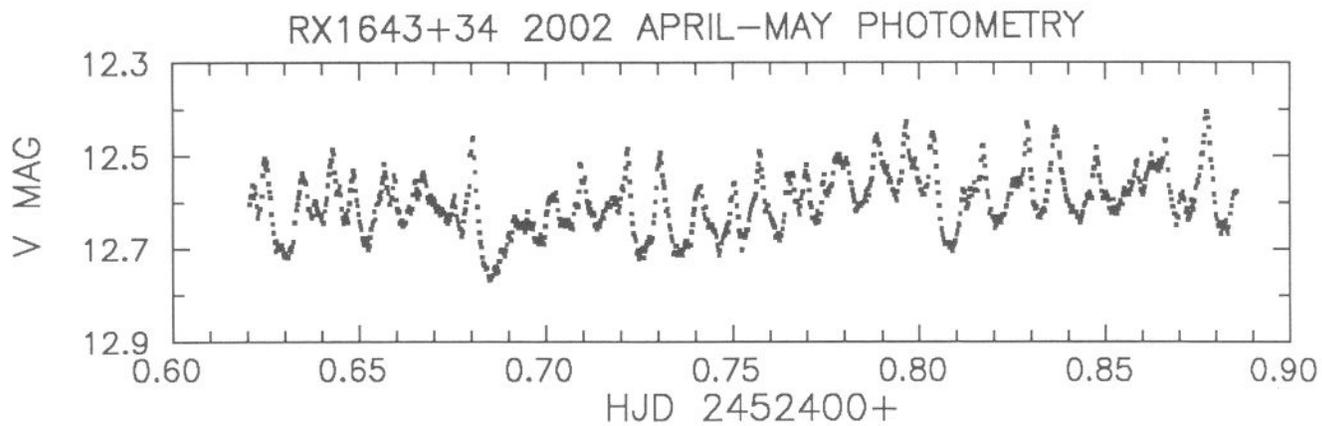

RX1643+34 2002 APRIL–MAY PHOTOMETRY

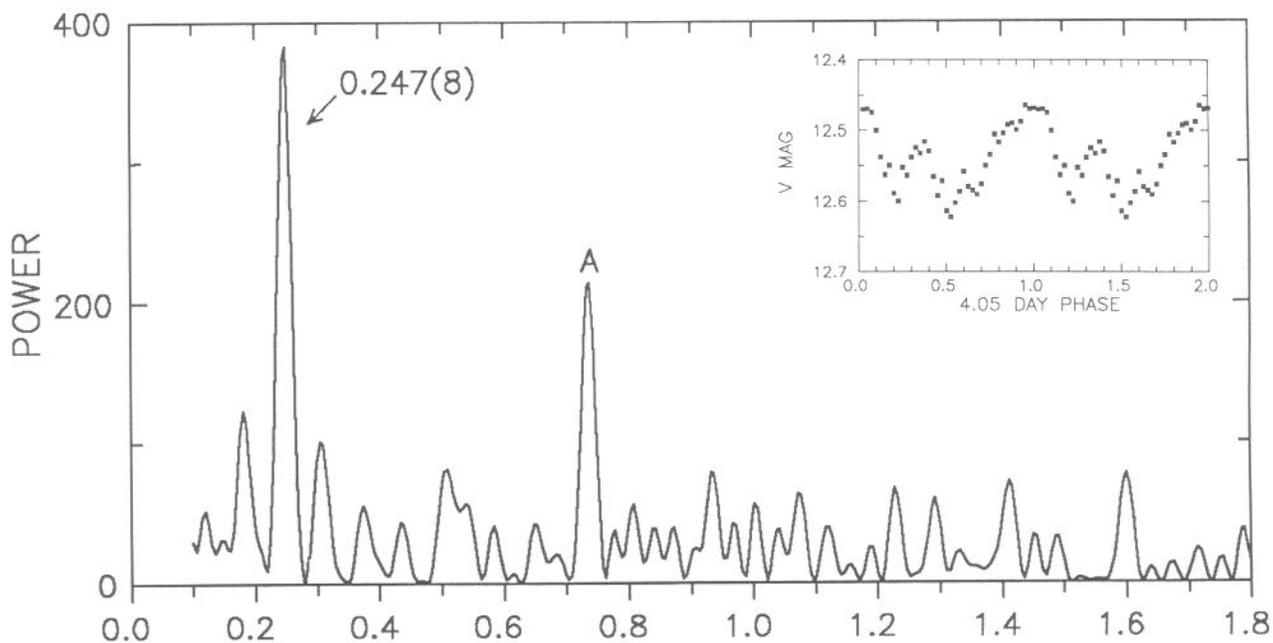

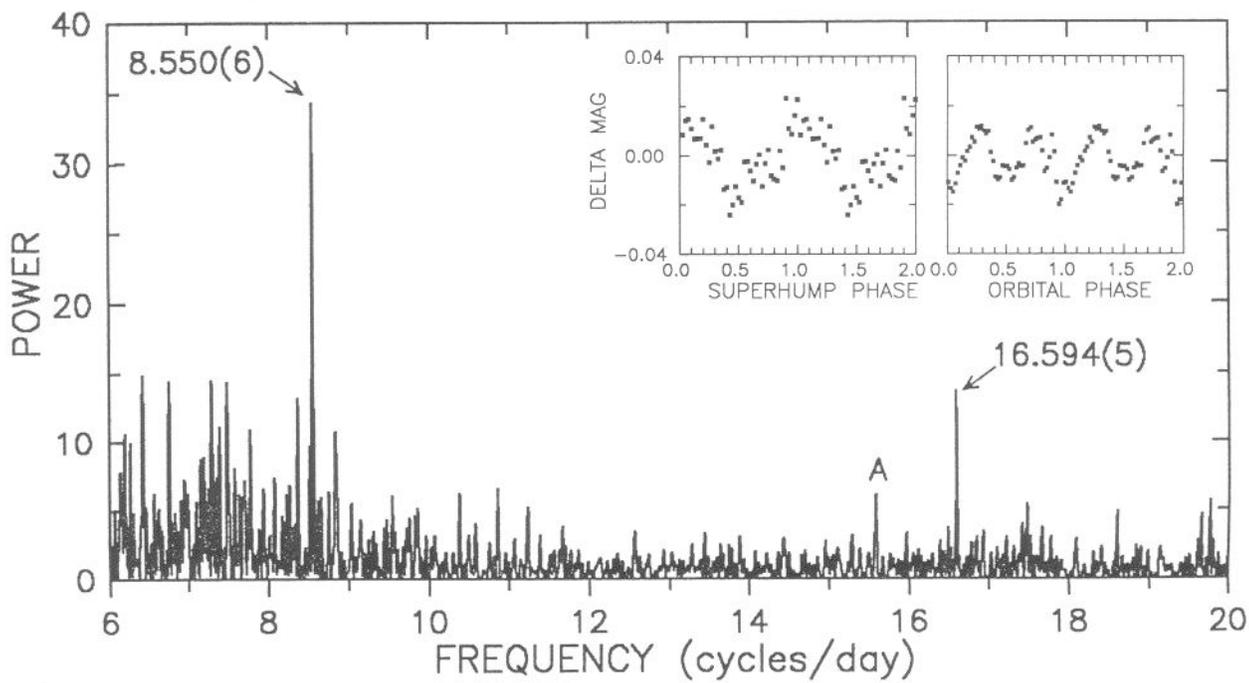

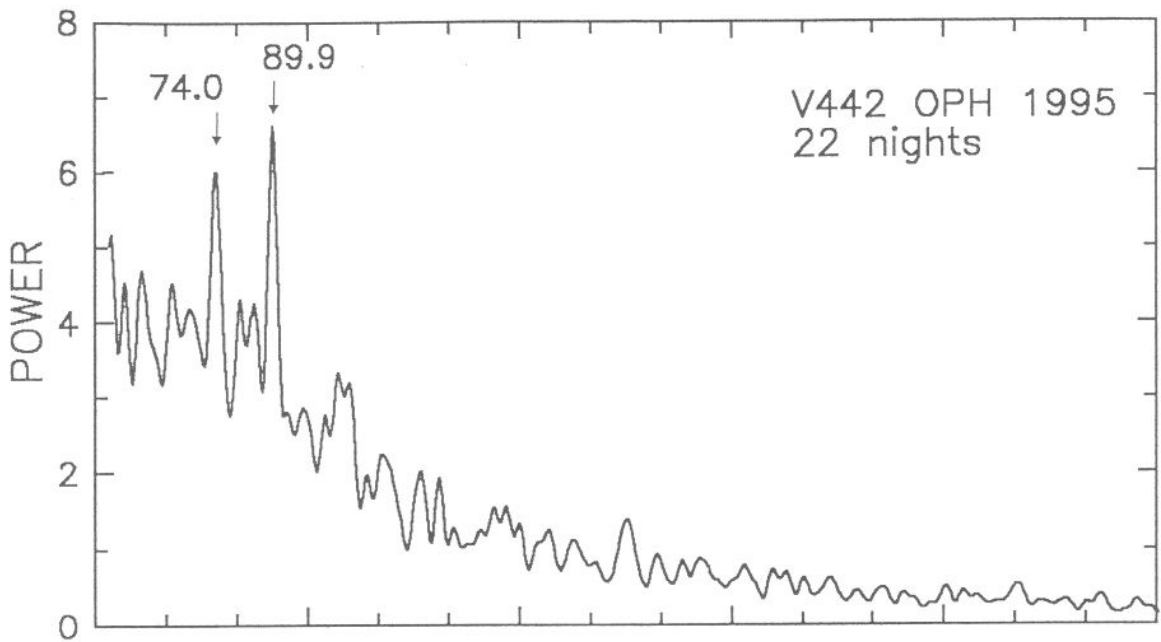

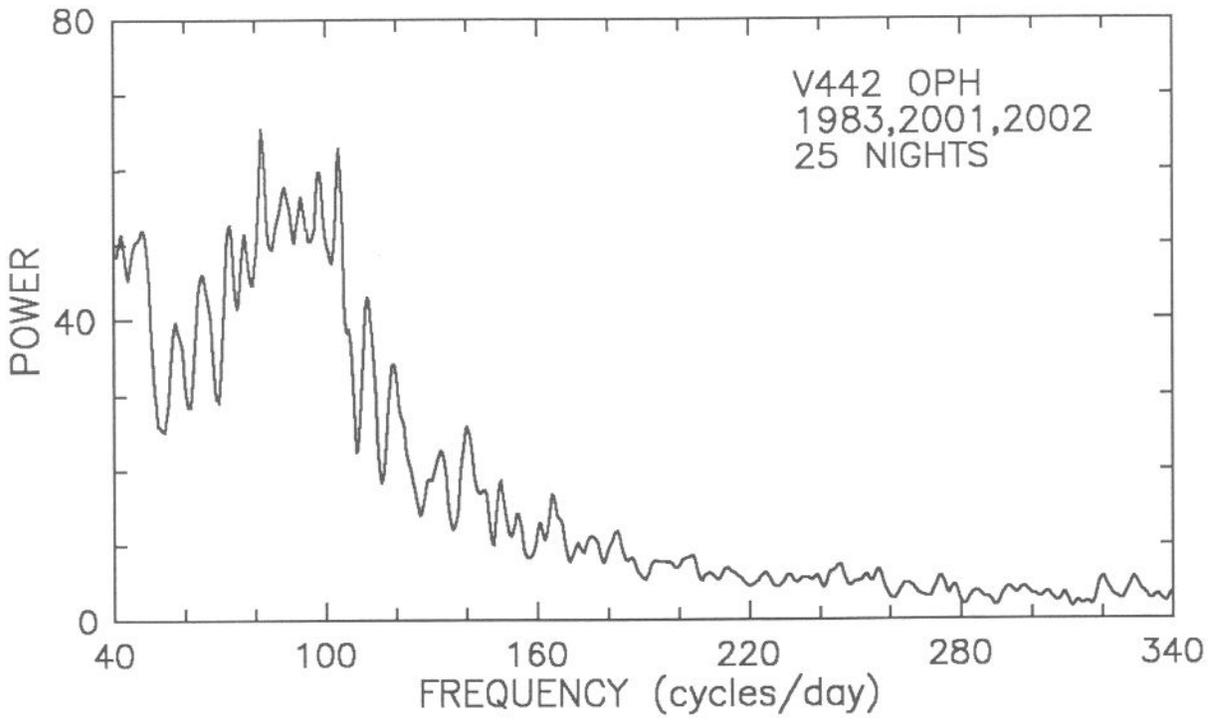

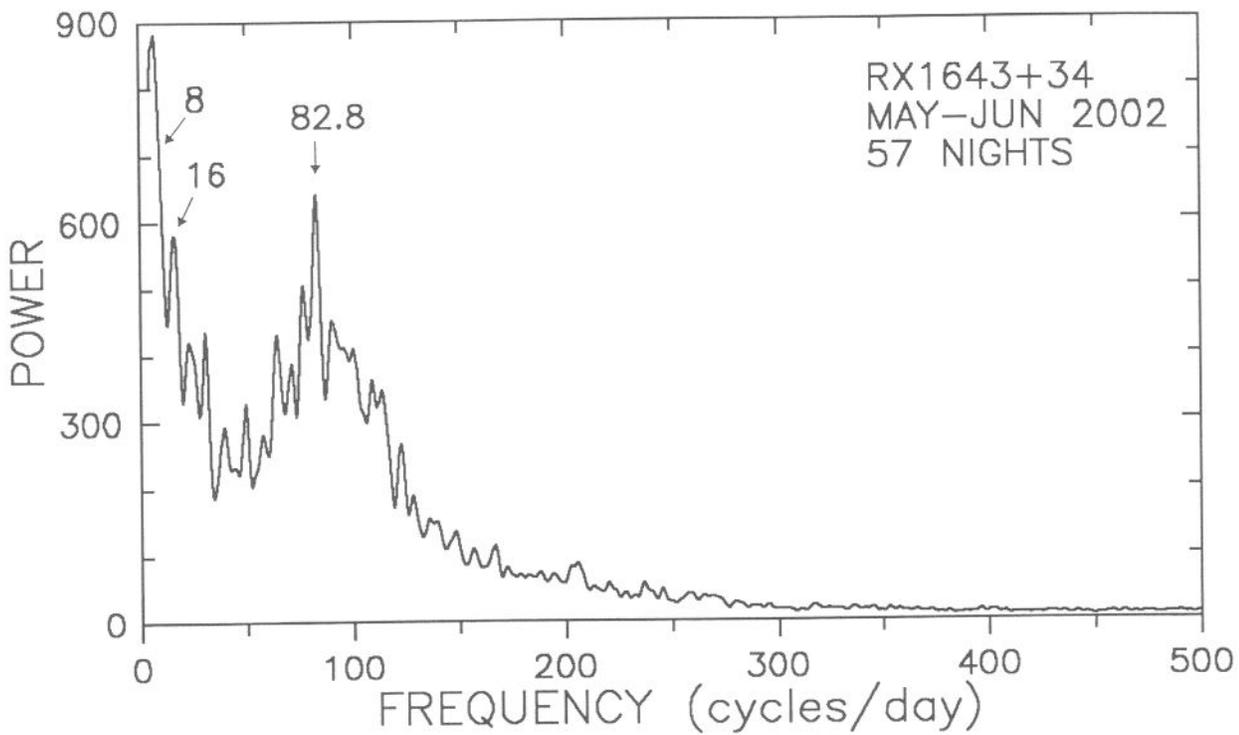

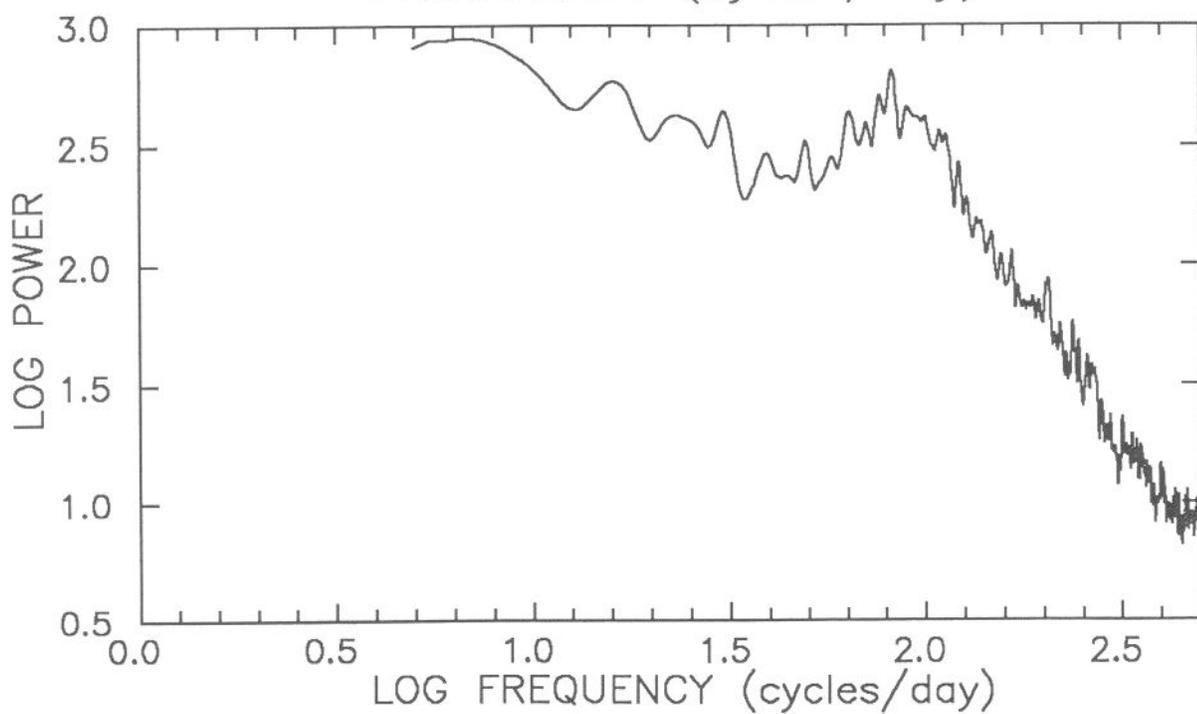

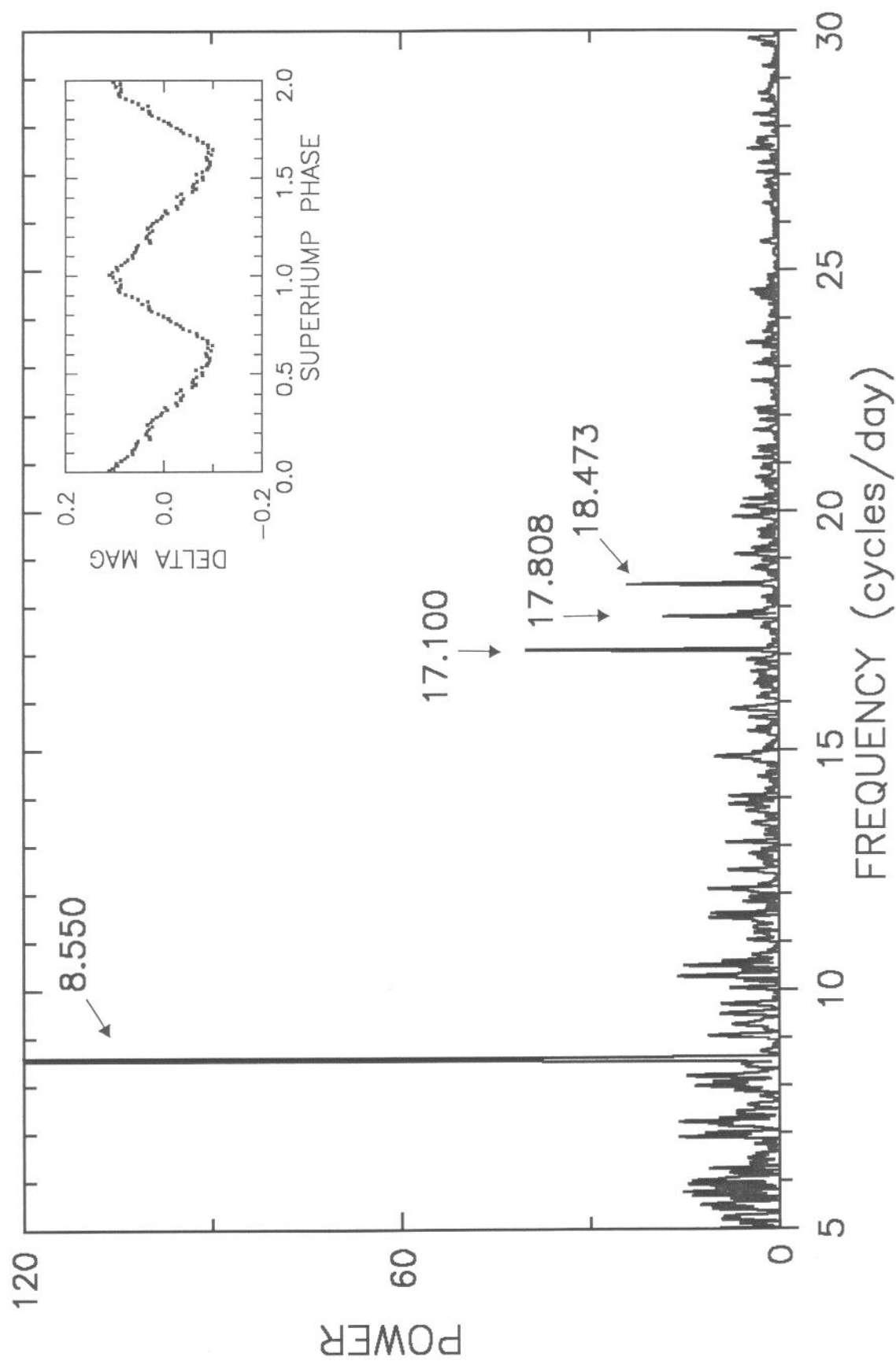

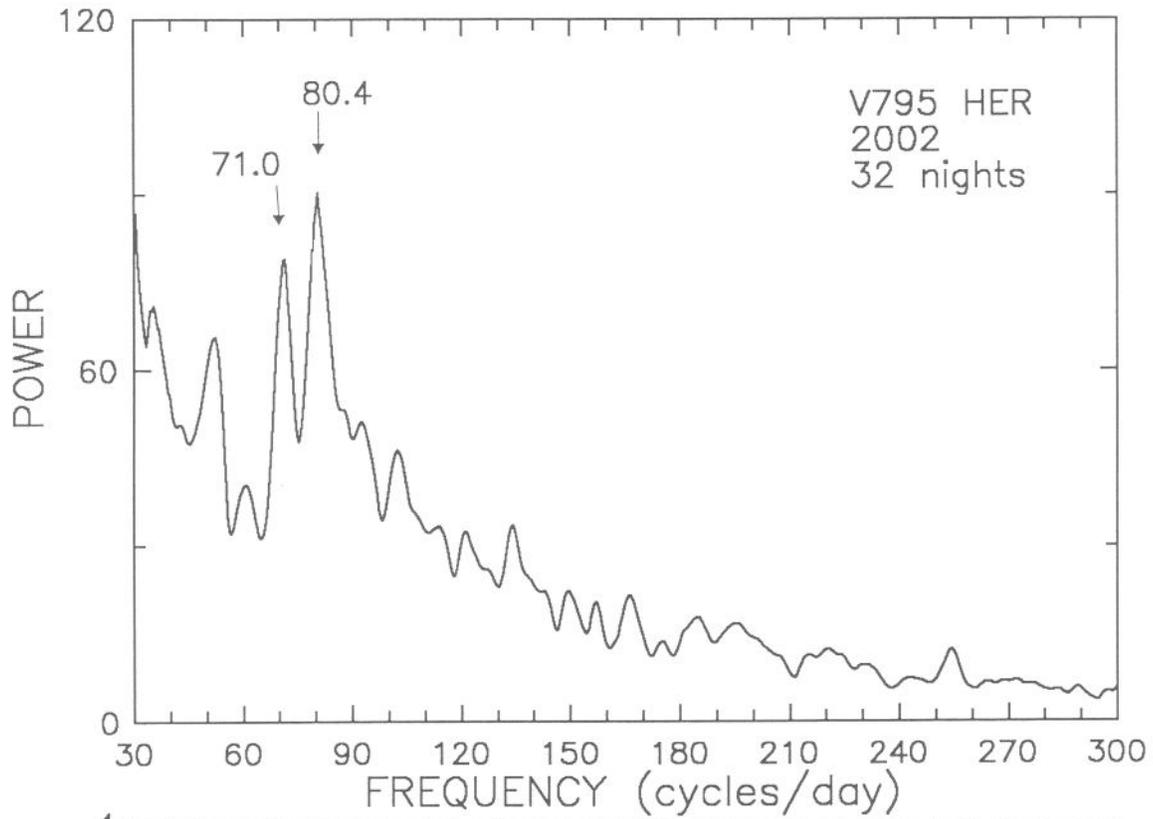

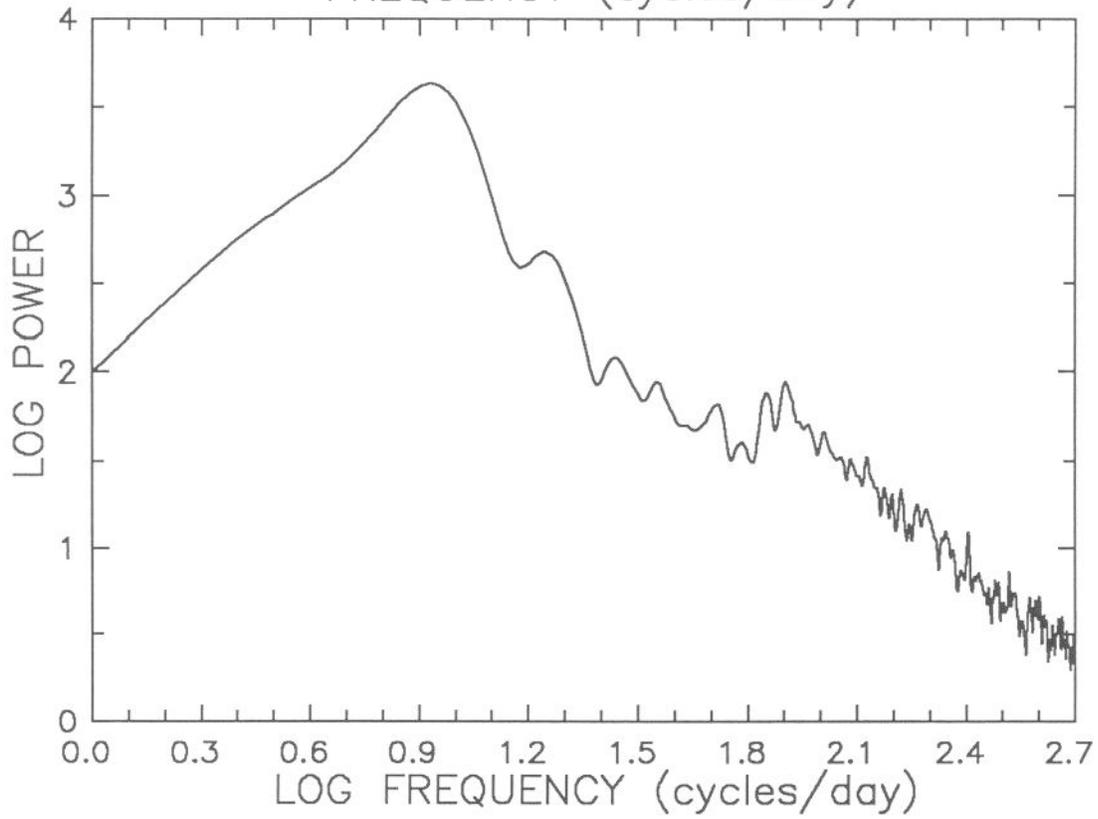

DW UMA SUPERHUMPS IN 2002 (POSITIVE) AND 1996 (NEGATIVE)

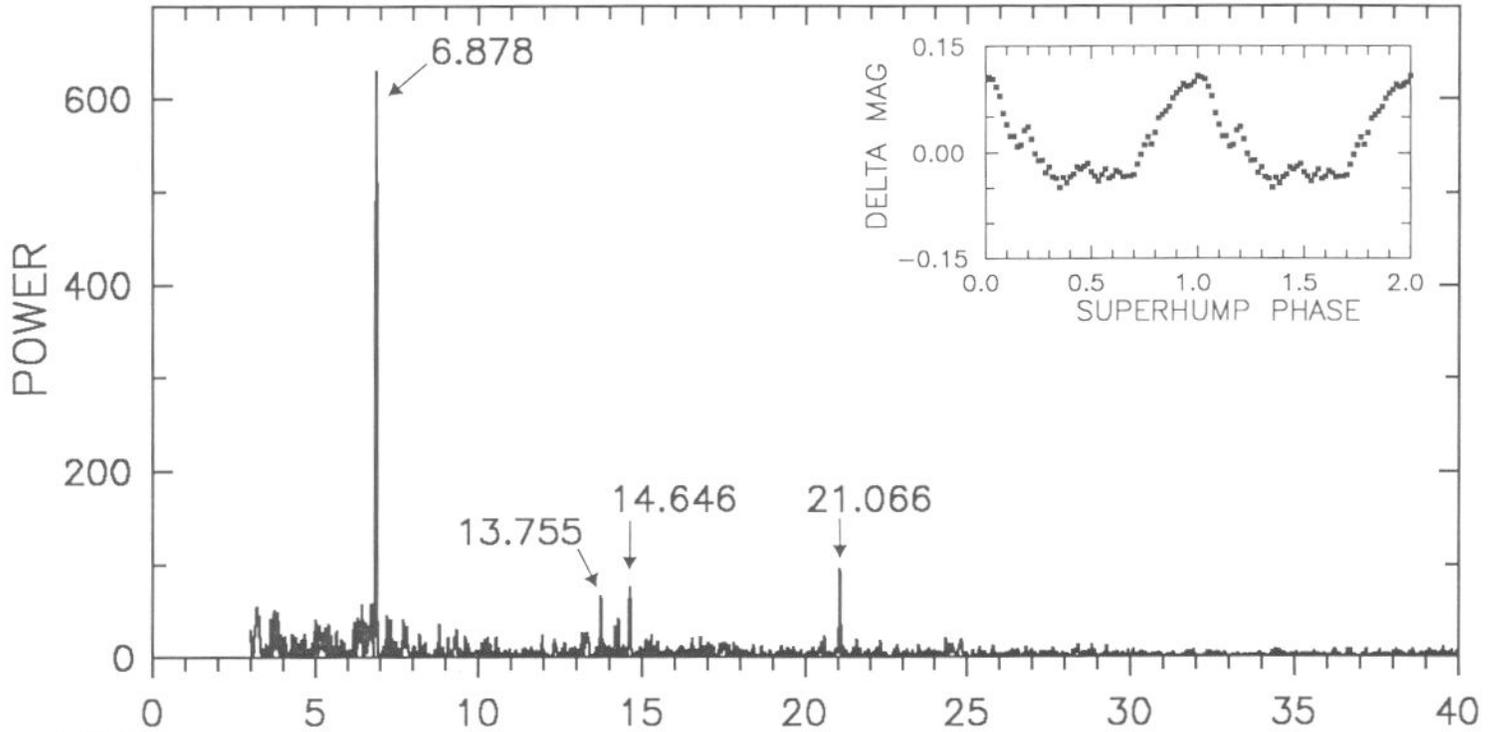

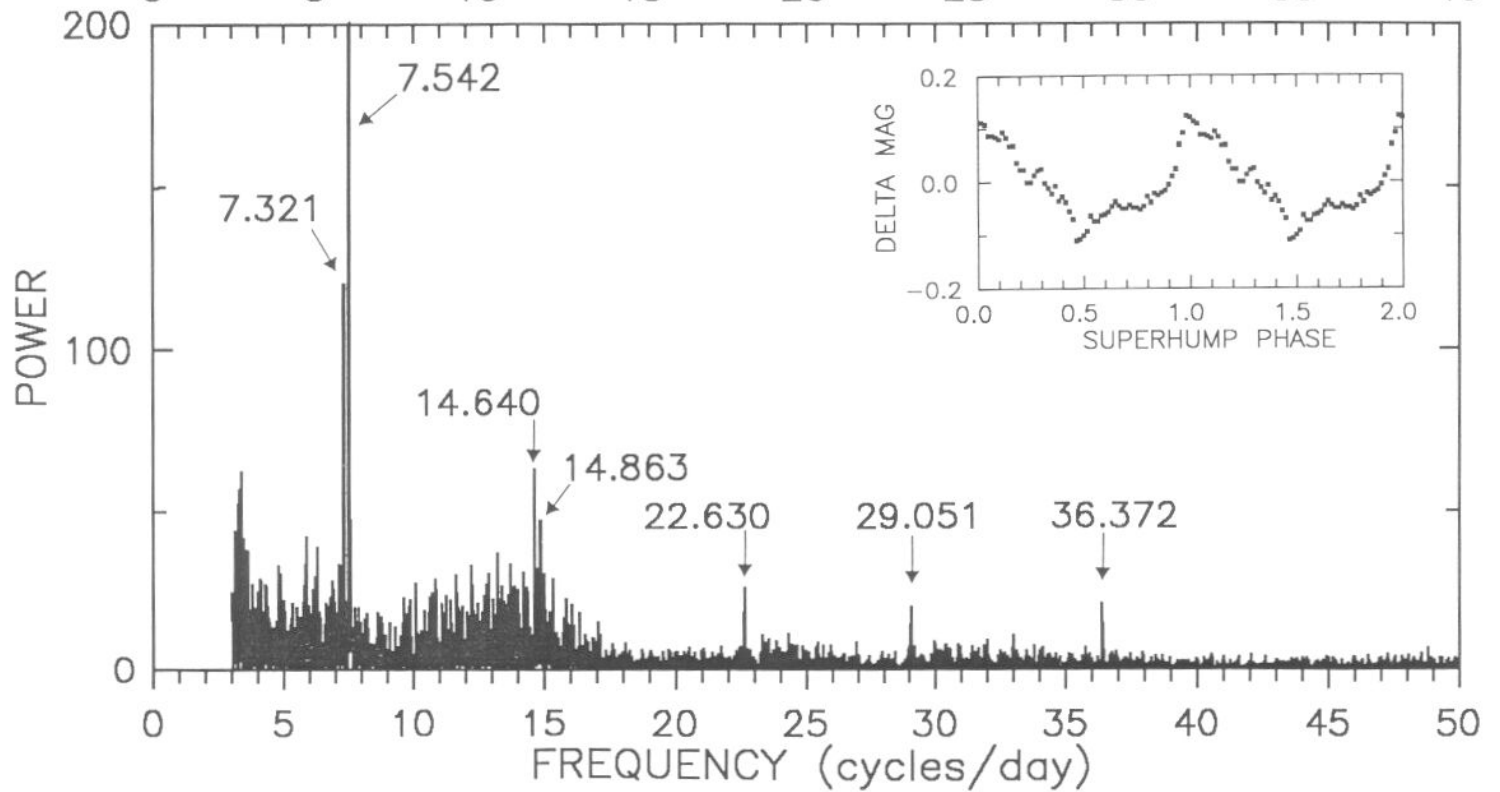

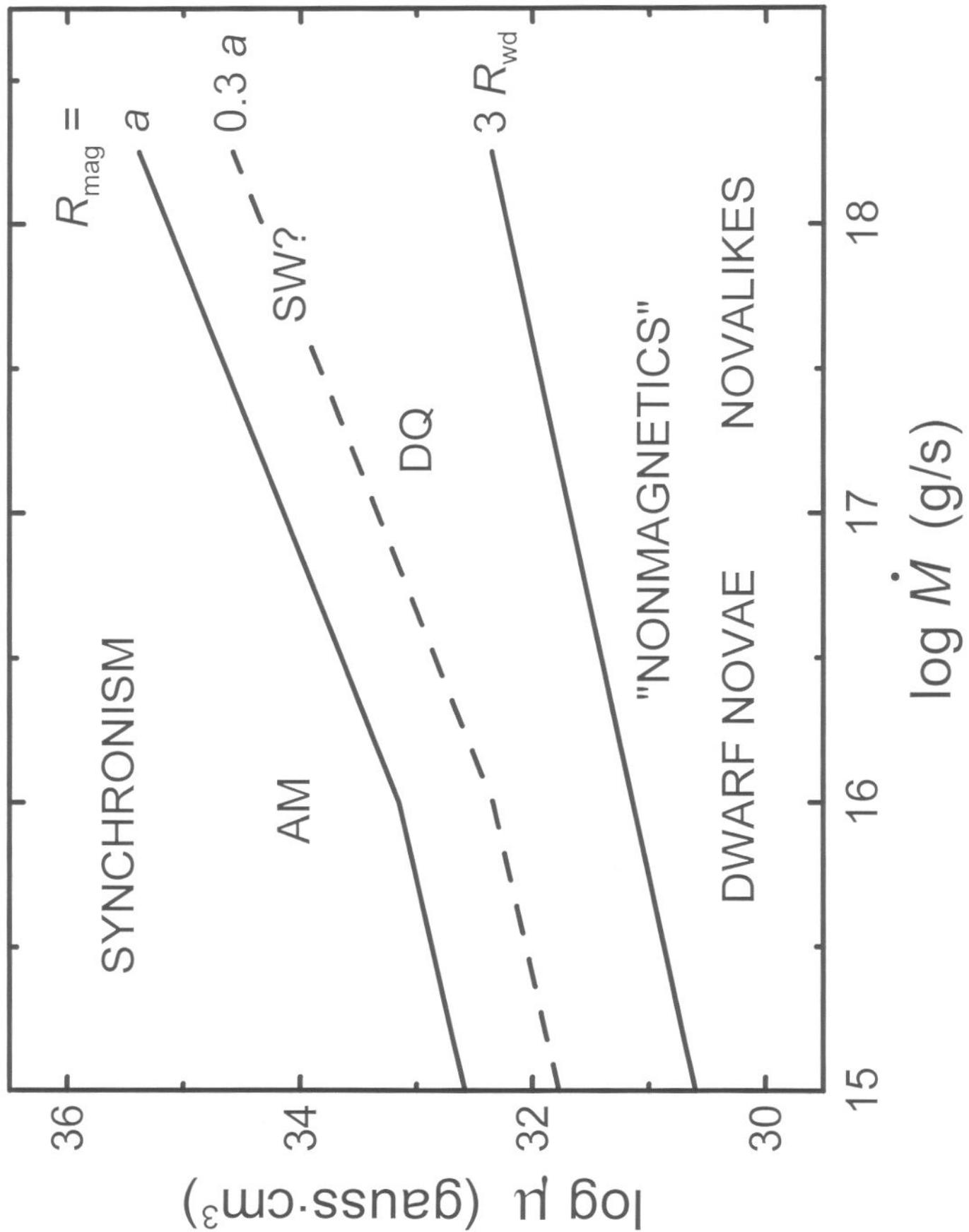